\documentclass[structabstract]{aa}  
\usepackage{graphicx}
\usepackage{txfonts}
\usepackage{natbib}
\begin{document}
\title{The high-mass disk candidates NGC7538IRS1 and
  NGC7538S\thanks{Based on observations carried out with the IRAM
    Plateau de Bure Interferometer. IRAM is supported by INSU/CNRS
    (France), MPG (Germany) and IGN (Spain). The data are available in
    electronic form at the CDS via anonymous ftp to
    cdsarc.u-strasbg.fr (130.79.128.5) or via
    http://cdsweb.u-strasbg.fr/cgi-bin/qcat?J/A+A/}.}


   \author{H.~Beuther
          \inst{1}
          \and
          H.~Linz
          \inst{1}
          \and
          Th.~Henning
           \inst{1}
           }
   \institute{$^1$ Max-Planck-Institute for Astronomy, K\"onigstuhl 17,
              69117 Heidelberg, Germany, \email{name@mpia.de}}



\abstract
{The nature of embedded accretion disks around forming high-mass
  stars is one of the missing puzzle pieces for a general
  understanding of the formation of the most massive and luminous
  stars.}
{We want to dissect the small-scale structure of the dust continuum
  and kinematic gas emission toward two of the most prominent
  high-mass disk candidates.}
{Using the Plateau de Bure Interferometer at $\sim$1.36\,mm
  wavelengths in its most extended configuration we probe the dust and
  gas emission at $\sim$0.3$''$, corresponding
  to linear resolution elements of $\sim$800\,AU.}
{Even at that high spatial resolution NGC7538IRS1 remains a single
  compact and massive gas core with extraordinarily high column
  densities, corresponding to visual extinctions on the order of
  $10^5$\,mag, and average densities within the central 2000\,AU of
  $\sim 2.1\times 10^9$\,cm$^{-3}$ that have not been measured before.
  We identify a velocity gradient across in northeast-southwest
  direction that is consistent with the mid-infrared emission, but we
  do not find a gradient that corresponds to the proposed CH$_3$OH
  maser disk. The spectral line data toward NGC7538IRS1 reveal strong
  blue- and red-shifted absorption toward the mm continuum peak
  position. While the blue-shifted absorption is consistent with an
  outflow along the line of sight, the red-shifted absorption allows
  us to estimate high infall rates on the order of
  $10^{-2}$\,M$_{\odot}$\,yr$^{-1}$.  Although we cannot prove that
  the gas will be accreted in the end, the data are consistent with
  ongoing star formation activity in a scaled-up low-mass star
  formation scenario. Compared to that, NGC7538S fragments in a
  hierarchical fashion into several sub-sources. While the kinematics
  of the main mm peak are dominated by the accompanying jet, we find
  rotational signatures from a secondary peak. Furthermore, strong
  spectral line differences exist between the sub-sources which is
  indicative of different evolutionary stages within the same
  large-scale gas clump.}
{NGC7538IRS1 is one of the most extreme high-mass disk candidates
  known today. The large concentration of mass into a small area
  combined with the high infall rates are unusual and likely allow
  continued accretion. While the absorption is interesting for the
  infall studies, higher-excited lines that do not suffer from the
  absorption are needed to better study the disk kinematics. In
  contrast to that, NGC7538S appears as a more typical high-mass star
  formation region that fragments into several sources. Many of them
  will form low- to intermediate-mass stars. The strongest mm
  continuum peak is likely capable to form a high-mass star, however,
  likely of lower mass than NGC7538IRS1.}  
   \maketitle

\section{Introduction}
\label{intro}

The characterization of accretion disks around young high-mass
protostars is one of the main unsolved questions in massive star
formation research (e.g.,
\citealt{beuther2006b,beuther2009c,cesaroni2007,kraus2010}).  The
controversy arises around the difficulty to accumulate mass onto a
massive protostar when it gets larger than 8\,M$_{\odot}$ because the
radiation pressure of the growing protostar may be strong enough to
revert the gas inflow in spherical accretion scenarios (e.g.,
\citealt{kahn1974,wolfire1987}).  Different ways to circumvent this
problem are proposed, the main two are (a) scaled-up disk accretion
(e.g., \citealt{yorke2002,krumholz2009,kuiper2010}) partially
requiring initial turbulent gas and dust cores (e.g.,
\citealt{mckee2003}) and including ionization radiation (e.g.,
\citealt{keto2002a}) and magnetic fields (e.g., \citealt{peters2011}),
and (b) competitive accretion and potential (proto)stellar mergers at
the dense centers of evolving massive (proto)clusters (e.g.,
\citealt{bonnell2004,bonnell2006,bally2005}).

Over recent years, much indirect evidence has been accumulated that
massive accretion disks do exist. The main argument stems from massive
molecular outflow observations that identify collimated and energetic
outflows from high-mass protostars, resembling the properties of known
low-mass star formation sites (e.g.,
\citealt{henning2000,beuther2002d,wu2004,zhang2005,arce2007,lopez2009}).
Such collimated jet-like outflow structures are only explainable with
an underlying massive accretion disk driving the outflows via
magneto-centrifugal acceleration. From a modeling approach, numeric
simulations and analytic calculations of massive collapsing gas cores
result in the formation of massive accretion disks as well
\citep{yorke2002,kratter2006,krumholz2009,kuiper2010,peters2011}.
Although alternative formation scenarios are proposed, there is a
growing consensus in the massive star formation community that
accretion disks should also exist in high-mass star formation.
However, it is still poorly known whether such massive disks are
similar to their low-mass counterparts, hence dominated by the central
protostar and in Keplerian rotation, or whether they are perhaps
self-gravitating non-Keplerian entities.

While indirect evidence for massive disks is steadily increasing,
direct observational studies are largely missing. This discrepancy can
mainly be attributed to the clustered mode of massive star formation,
the typically large distances and the high extinction.  Hence,
spatially disentangling such structures is a difficult task. While
several disk candidates exist around early B-stars (e.g.,
\citealt{cesaroni1997,cesaroni2005,schreyer2002,shepherd2001,zhang2002,chini2004,kraus2010,keto2010,fallscheer2011}),
more massive O-star like systems rather show larger-scale toroid-like
structures not consistent with classical Keplerian accretion disks
(e.g.,
\citealt{beltran2004,beltran2011,beuther2005c,beuther2008a,sollins2005,keto2006,cesaroni2007,fallscheer2009}).
However, the non-detection of Keplerian structures around O-stars does
not imply that they do not exist, it rather indicates that they are
likely on smaller spatial scales hidden by the toroids. Therefore,
penetrating more deeply into the central structures at the highest
possible spatial resolution is the next step to go.

\paragraph{The high-mass accretion disk candidates
  NGC7538IRS1 and NGC7538S:}

\begin{figure}[htb] 
\includegraphics[width=9cm]{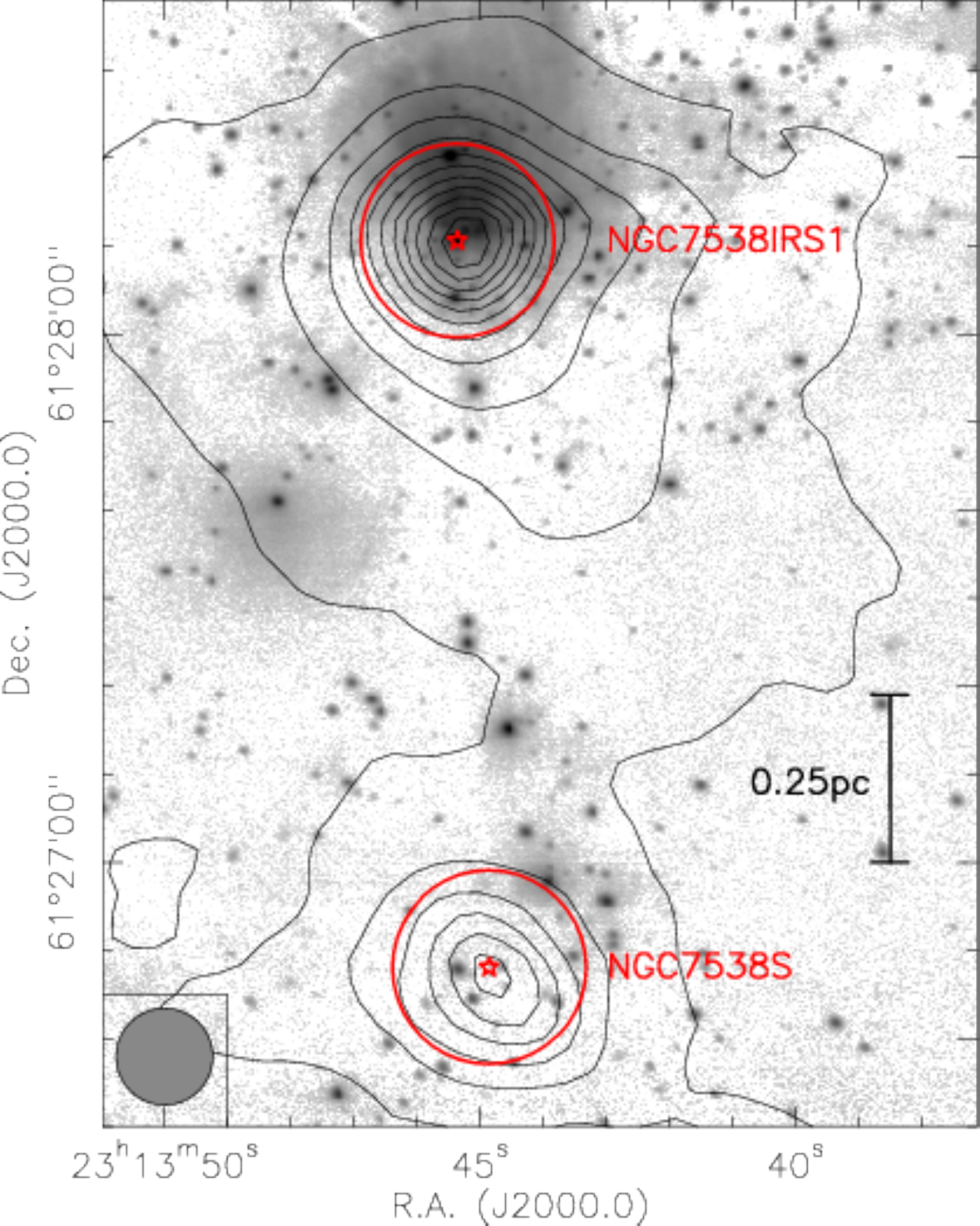}
\caption{Overview of the NGC7538 complex. The grey-scale presents the
  K-band image from \citet{puga2010}, and the contours show the
  single-dish 1.2\,mm continuum data from \citet{sandell2004}. The
  colored stars and circles show the central positions and FWHM of the
  primary beam of the PdBI at 1.3\,mm wavelength for NGC7538IRS1 and
  NGC7538S in the north and south, respectively. The contour levels
  are from 250\,mJy\,beam$^{-1}$ to 5.25\,Jy\,beam$^{-1}$ in steps of
  500\,mJy\,beam$^{-1}$. The beam of the 30\,m observations and a
  scale-bar are shown at the bottom-left and right, respectively.}
\label{overview}
\end{figure}

The source selection for such a project is driven by the scientific
goals and the technical feasibility. The two disk candidates
NGC7538IRS1 and NGC7538S combine the best of both worlds: On the one
hand, they are two already well-studied massive accretion disk
candidates in different evolutionary stages at a still modest distance
of $\sim$2.7\,kpc (e.g.,
\citealt{sandell2003,sandell2010,pestalozzi2004,pestalozzi2006,moscadelli2009,puga2010}),
and they are easily observable in a mosaic mode since their spatial
separation is only of order $\sim$80$''$. On the other hand, with a
R.A.~of 23 hours and a Decl.~of 61\,degrees they are ideal Plateau de
Bure Interferometer (PdBI) targets to be observed in long tracks
resulting in the best achievable synthesized beams. Figure
\ref{overview} presents a large-scale 1.2\,mm continuum map on top of
near-infrared K-band data \citep{sandell2004,puga2010} (see also
\citealt{reid2005}).

\begin{figure*}[ht!] 
\includegraphics[width=18.4cm]{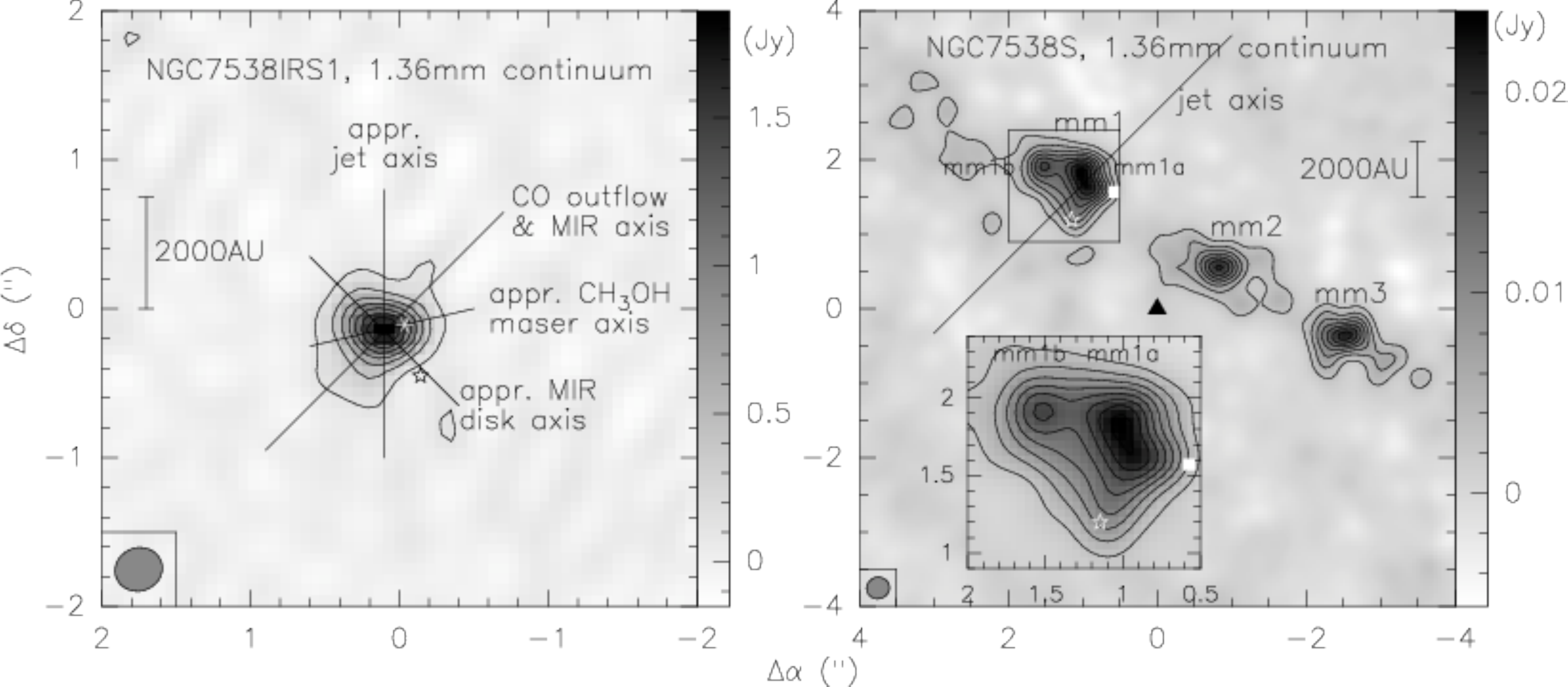}
\caption{PdBI 1.36\,mm continuum images toward NGC7538IRS1 and
  NGC7538S in the left and right panel, respectively. The contour
  levels start at $4\sigma$ values and continue in $8\sigma$ and
  $4\sigma$ steps for NGC7538IRS1 and NGC7538S ($1\sigma$ values are
  29\,mJy\,beam$^{-1}$ and 0.7\,mJy\,beam$^{-1}$, respectively).
  Several potential disk and outflow axis reported in the literature
  are presented
  (\citealt{davis1998,debuizer2005c,sandell2009,pestalozzi2004,pestalozzi2009,sandell2010},
  see Introduction for more details).  The open stars, square,
  triangle and six-pointed star mark the positions of the OH, H$_2$O,
  class {\sc ii} CH$_3$OH and H$_2$CO masers
  \citep{argon2000,kameya1990,pestalozzi2006,hoffman2003}. A scale-bar
  and the synthesized beam are shown in both panels. The box zooms
  into the region around mm1 in more detail.  The coordinates are
  relative to the phase centers given in sec.~\ref{obs}.}
\label{cont}
\end{figure*}

{\bf NGC7538IRS1} has been extensively studied, and the central source
is estimated to a mass of $\sim$30\,M$_{\odot}$ and a luminosity of
$\sim 8\times 10^4$\,L$_{\odot}$ (e.g.,
\citealt{willner1976,campbell1984,pestalozzi2004}). While
\citet{campbell1984b} and \citet{sandell2009} report an ionized jet in
north-south direction, \citet{minier2000} and
\citet{pestalozzi2004,pestalozzi2009} present CH$_3$OH maser
observations indicative of an accretion disk almost perpendicular to
the outflow. Partly different interpretations arise from mid-infrared
continuum imaging \citep{debuizer2005c}: They detected elongated
mid-infrared emission in northwest-southeast direction aligned with a
bipolar outflow reported in CO by \citet{davis1998}, and even earlier
in NH$_3$ by \citet{keto1991} who also performed radiation transfer
calculations for the region. This large-scale mid-infrared emission
may stem from the inner walls of the outflow.  The jet and outflow
emission is interpreted from speckle data as due to a precessing jet
\citep{kraus2006}. On smaller scales, the mid-infrared emission
appears elongated almost perpendicular to that outflow axis which
\citet{debuizer2005c} interprete as an inner accretion disk of
approximate size of $\sim 900$\,AU.  \citet{klaassen2009} also
identify a velocity gradient in dense gas in northeast-southwest
direction. Furthermore, the rare H$_2$CO maser emission from this
source \citep{forster1985,pratap1992,hoffman2003} is also consistent
with a very young disk candidate.  \citet{surcis2011} present new
CH$_3$OH and H$_2$O maser observations, and their data are also
consistent with a rotating structure in northeast-southwest direction
and an outflow opposite to that.  \citet{hutawarakorn2003} show the OH
maser emission in this region.  Figure \ref{cont} sketches the
different axis and other features found in the literature. While
maser, ionized gas and warm dust are well studied for this source, a
good characterization of the dust and thermal gas emission was
lacking.  Recently, \citet{qiu2011} observed the region with the
Submillimeter Array (SMA) at $3''\times 2''$ resolution in mm
continuum and line emission, and they revealed 9 mm sources within a
projected area of 0.35\,pc. Compared to the blue-shifted absorption
observed by \citet{keto1991} and \citet{zheng2001} in the
low-excitation NH$_3$ lines that is indicative of outflowing gas
motions, \citet{qiu2011} detected first red-shifted absorption toward
the main mm core which they interprete as signature of ongoing infall.
Furthermore, the general structure of their proposed multiple outflows
is consistent with the northwest-southeast outflow previously reported
by, e.g., \citet{keto1991,davis1998,klaassen2011}. Our data now
resolve the region at again an order of magnitude higher spatial
resolution, allowing us to study the central core in unprecedented
detail.

{\bf NGC7538S} is supposed to be younger than NGC7538IRS1 but also
hosts CH$_3$OH Class {\sc ii}, H$_2$O and OH maser emission
\citep{kameya1990,argon2000,pestalozzi2006}.  About $\sim 80''$ south
of NGC7538IRS1 (Fig.~\ref{overview}), it roughly coincides with a
far-infrared source with luminosity $\sim 1.5\times 10^4$\,L$_{\odot}$
\citep{werner1979,thronson1979} corresponding to an early B-star.
Recently, \citet{sandell2010} and \citet{wright2012} also report the
detection of NGC7538S in deep mid-infrared IRAC Spitzer observations
at wavelengths between 4.5 and 8\,$\mu$m, as well as a weak cm
continuum source likely stemming from a thermal jet.
\citet{sandell2003} resolved a 30000\,AU rotating structure with a
bipolar outflow emanating perpendicular to that by means of BIMA
interferometric mm observations . \citet{wright2012} resolved that
structure into three separate mm sources. Recent interferometric
observations (between $\sim 3''$ and $8''$ resolution) in different
molecular line tracers largely confirm this picture
\citep{sandell2010}. However, their work indicated that many of the
molecules are affected by the jet/outflow, and clear rotational
signatures were hard to isolate. While the previous data are
consistent with a rotating structure, the spatial resolution was not
sufficient to analyze the disk candidate in detail.  Figure \ref{cont}
again sketches the main features found in the literature.

The different ages of the two targets make them ideal candidates
to investigate also disk evolutionary properties
within the same observations.

\section{Observations} 
\label{obs}

The two sources were observed in two tracks -- A and B configuration
on January 26th, 2011, and February 10th, 2011, respectively -- in two
fields where each field was centered on NGC7538IRS1 and
NGC7538S (see Fig.~\ref{overview}). The phase centers for the two
fields were R.A (J2000.)  23h\,13m\,45.36s, Dec.~(J2000.0)
61$^o$\,28$'$\,10.55$''$ and R.A.~(J2000.) 23h\,13m\,44.86s and
Dec.~(J2000.0) 61$^o$\,26$'$\,48.10$''$. Phase calibration was
conducted with regularly interleaved observations of the quasars
2146+608, 0059+581 and 0016+731. The bandpass and flux were calibrated
with observations of 3C345, 3C273 and MWC349. The absolute flux level
is estimated to 20\% accuracy.

The continuum emission was extracted from broad band data obtained
with the WIDEX correlator with four units and two polarizations
covering the frequency range from 217.167 to 220.836\,GHz. NGC7538IRS1
is such an extremely line-rich source that barely any line-free region
exists in the spectrum. Therefore, extracting a line-free continuum is
difficult, and we produced our continuum map from the whole bandpass.
Although there obviously is some line contamination in the final
continuum image, the continuum level itself ($>1.8$\,Jy\,beam$^{-1}$)
is so extraordinarily high that the relative contribution from the
lines is negligible. To get a quantitative estimate of the line
contamination, we also produced a continuum image from only a very
small line-free region ($\sim 71\,MHz$) between the CH$_3$CN lines.
The peak flux in this image is only $1.1$\% below the peak flux in the
continuum map produced from the whole bandpass. Obviously the line
contamination is negligible and we use the map based on the full
bandpass because of the lower rms noise.  NGC7538S is less line-rich,
and therefore, we could produce the continuum image from the line-free
parts of the spectrum.  A comparison of the continuum images for
NGC7538S with and without line contamination shows that they are
almost identical. Although NGC7538IRS1 exhibits more line emission,
the previous comparison also indicates that including the lines in the
NGC7538IRS1 continuum bandpath only marginally changes the real
fluxes.  The full bandpath spectral line data with a chemical analysis
will be presented in a forthcoming paper. The $1\sigma$ continuum rms
for NGC7538IRS1 and NGC7538S are 29\,mJy\,beam$^{-1}$ and
0.7\,mJy\,beam$^{-1}$. The difference in rms can be explained by the
fact that for sparse antenna interferometers like the PdBI, the rms is
usually not the thermal rms but it is dominated by the side-lobes of
the strongest source in the field. And since NGC7538IRS1 is far
brighter than NGC7538S, also the rms for NGC7538I is significantly
higher.

\begin{table}[htb]
\caption{Observed spectral lines}
\begin{tabular}{lrr}
\hline \hline
Freq. & Mol. & $E_u/k$ \\
(GHz) &       &  (K) \\
\hline
218.222 & H$_2$CO$(3_{0,3}-2_{0,1})$ & 21 \\
218.298 & HCOOCH$_3(17_{3,14}-16_{3,13})$ & 100  \\
218.325 & HC$_3$N$(24-23)$ & 131 \\
218.440 & CH$_3$OH$(4_{2,2}-3_{1,2})$ & 46 \\
218.460 & NH$_2$CHO$(10_{1,9}-9_{1,8})$ & 61 \\
218.476 & H$_2$CO$(3_{2,2}-2_{2,1})$ & 68 \\
218.903 & OCS$(18-17)$ & 100 \\
220.167 & HCOOCH$_3(17_{4,13}-16_{4,12})$ & 103 \\
220.178 & CH$_2$CO$(11_{1,11}-10_{1,10})$ & 77 \\
220.190 & HCOOCH$_3(17_{4,13}-16_{4,12})$ & 103  \\
220.594 & CH$_3$CN$(12_6-11_6)$ & 326 \\
220.641 & CH$_3$CN$(12_5-11_5)$ & 248 \\
220.679 & CH$_3$CN$(12_4-11_4)$ & 183 \\
220.709 & CH$_3$CN$(12_3-11_3)$ & 133 \\
220.730 & CH$_3$CN$(12_2-11_2)$ & 98 \\
220.743 & CH$_3$CN$(12_1-11_1)$ & 76 \\
220.747 & CH$_3$CN$(12_0-11_0)$ & 69 \\
\hline \hline
\end{tabular}
~\\
\label{linelist}
\end{table}

To extract kinematic information, we put several high-spectral
resolution units with a nominal resolution of 0.312\,MHz or
0.42\,km\,s$^{-1}$ into the bandpass covering the spectral lines
listed in Table \ref{linelist}.  The spectral line rms for
0.5\,km\,s$^{-1}$ wide spectral channels measured in emission-free
channels is 8\,mJy\,beam$^{-1}$ and 7\,mJy\,beam$^{-1}$ for
NGC7538IRS1 and NGC7538S, respectively.  The $v_{\rm{lsr}}$ for
NGC7538IRS1 and NGC7538S are $\sim -57.3$\,km\,s$^{-1}$ and $\sim
-56.4$\,km\,s${-1}$, respectively (Gerner et al.~in prep.,
\citealt{vandertak2000,sandell2010}).
The data were inverted with a ``robust'' weighting scheme and cleaned
with the clark algorithm. The synthesized beam of the final continuum
and line data is $\sim 0.31''\times 0.29''$ (PA 110$^{\circ}$).

\begin{figure}[htb] 
\includegraphics[width=9cm]{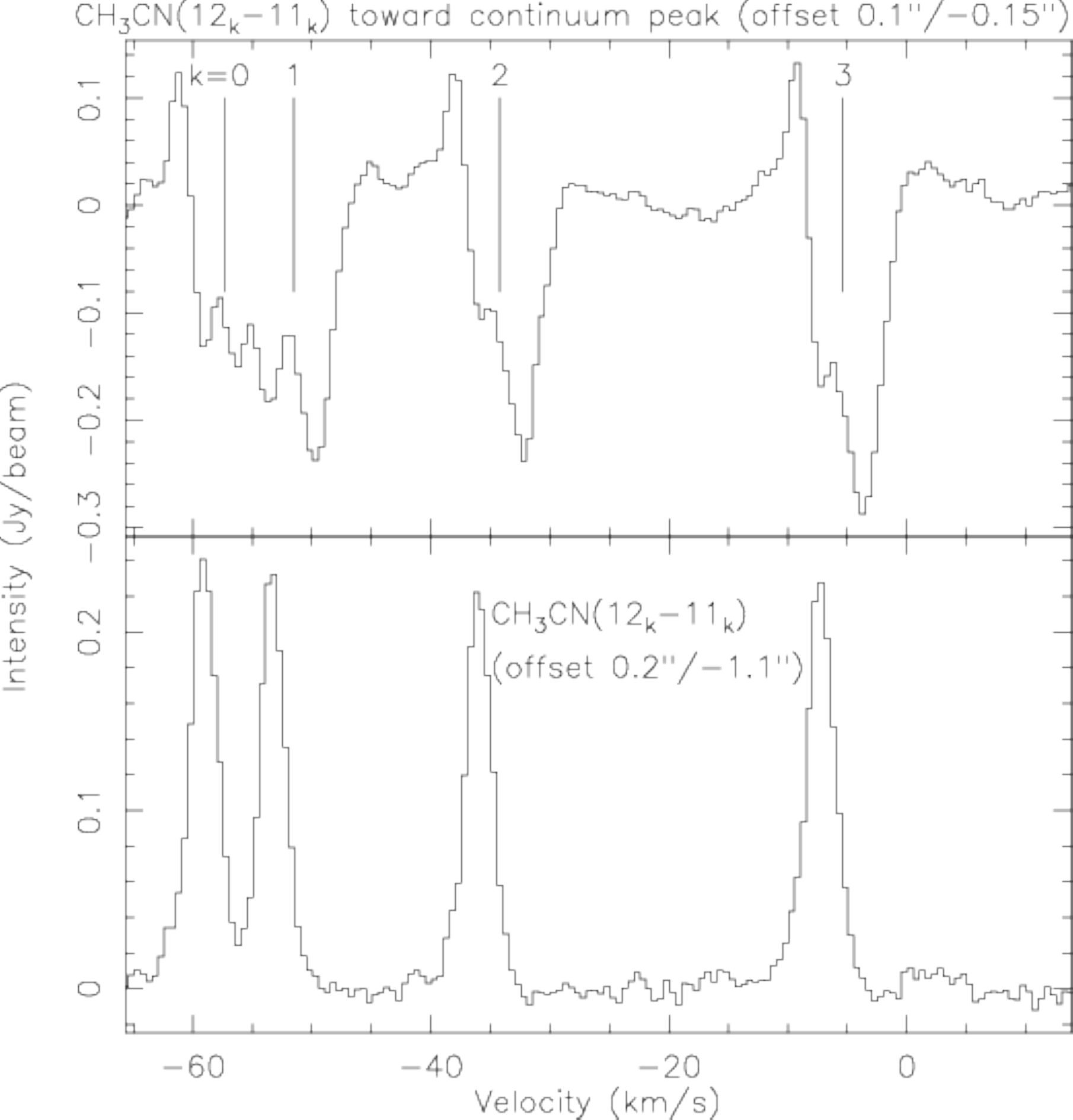}
\caption{CH$_3$CN$(12_k-11_k)$ spectra for $k=0...3$ toward the
  1.36\,mm continuum peak (top panel) and a reference position offset
  by $0.2''/-1.1''$ from NGC7538IRS1. The velocity of rest is always
  marked with a vertical line.}
\label{ch3cn_spectra_irs1}
\end{figure}

\begin{figure*}[htb] 
\includegraphics[width=18.4cm]{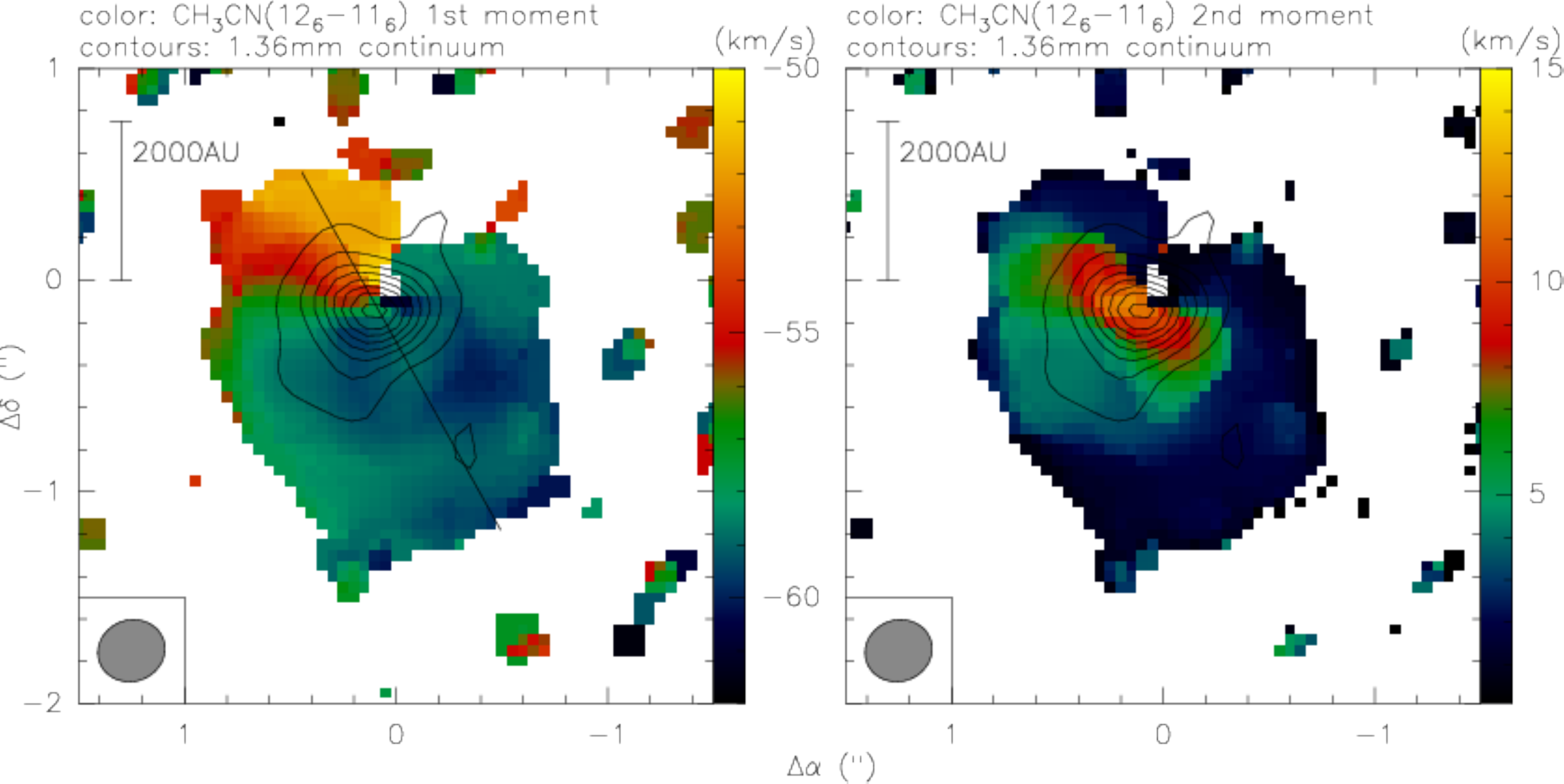}
\caption{The left and right panel present in color the 1st and 2nd
  moment maps in CH$_3$CN$(12_6-11_6)$ toward NGC7538IRS1. The
  contours show the 1.36\,mm continuum emission with the same contour
  levels as in Fig.~\ref{cont}. A scale-bar and the synthesized beam
  are presented in each panel. The line in the left panel outlines the
  axis for the pv-cuts in Figures \ref{irs1_pv} and \ref{irs1_pv2}.}
\label{ch3cn_mom}
\end{figure*}

\section{Results}

\subsection{Continuum emission}
\label{continuum}

Zooming in from the large-scale emission (Fig.~\ref{overview}), Figure
\ref{cont} presents the small-scale structure of the region in the
1.36\,mm continuum emission at a spatial resolution of $\sim 0.3''$ or
$\sim 800$\,AU. While NGC7538IRS1 remains a single source with peak
flux in excess of 1.8\,Jy\,beam$^{-1}$, NGC7538S is resolved into
several sub-sources labeled mm1 to mm3. NGC7538S mm1 shows even
additional fragmented substructure which we label as mm1a and mm1b.
The sub-source mm1a is elongated in approximately the north-south
direction, and it is likely that this elongation corresponds to an
unresolved substructure again. Comparing the high-resolution mm
continuum data with the near-infrared image by \citet{puga2010}, the
mm peak in NGC7538IRS1 is clearly associated with the main infrared
source IRS1, whereas the three mm peaks in NGC7538S have no near-infrared
counterpart. However, NGC7538S mm1 has recently been detected by
Spitzer at wavelengths between 4.5 and 8\,$\mu$m
\citep{sandell2010,wright2012}. It should also be noted that the 8
additional sources reported by \citet{qiu2011} within an area of
0.35\,pc are not detected by our higher-resolution PdBI observations.
This differences can be attributed to our smaller primary beam (FWHM
of $\sim 22''$) as well as the lower brightness sensitivity one
automatically achieves when going to higher spatial resolution (our
$3\sigma$ continuum rms of 87\,mJy\,beam$^{-1}$ corresponds to an
approximate brightness sensitivity of $\sim$24\,K).

Table \ref{mm} presents the measured peak and integrated fluxes of the
sub-sources shown in Fig.~\ref{cont}. The integrated fluxes are
measured within the $4\sigma$ contours. The single-dish 1.2\,mm MAMBO
data shown in Figure \ref{overview} exhibit peak fluxes of 5702 and
2872\,mJy\,beam$^{-1}$ for NGC7538IRS1 and NGC7538S, respectively.
Comparing these numbers to the integrated fluxes we measure with the
PdBI (Table \ref{mm}), we find that toward NGC7538IRS1 only $\sim
48$\% of the flux is filtered out with the interferometer.
\citet{qiu2011} also measure with the SMA at about an order of
magnitude lower spatial resolution ($3''\times 2''$) an integrated
flux of 3.6\,Jy, only $\sim$20\% higher than our fluxes measured with
$\sim 0.3''$ resolution. In comparison to that, toward NGC7538S
approximately 90\% of the single-dish flux is missing in the PdBI
data. This large difference indicates that NGC7538IRS1 is extremely
concentrated toward the central mm continuum peak whereas NGC7538S
exhibits emission on much larger scales. Since NGC7538IRS1 has
significant amounts of free-free emission (e.g.,
\citealt{pratap1992,keto2008,sandell2009}), we correct the fluxes for
that contribution in Table \ref{mm}. As shown by \citet{keto2008},
several H{\sc ii} region models can fit the data, and the exact
free-free contribution is hard to isolate. Here, we assume
$\sim$1000\,mJy to be produced by the free-free emission, the rest is
attributed to the dust emission.  Recently, \citet{wright2012} report
the cm free-free fluxes from NGC7538S mm1a, and following their
approach we correct 8\,mJy free-free flux contribution in Table
\ref{mm} as well.

\begin{table}[htb]
\caption{Millimeter continuum properties}
\begin{tabular}{lrr|rr|rr}
\hline \hline
Source & $S_{\rm{peak}}$     & $S_{\rm{int}}^e$ & $M$          & $N$ & $M$          & $N$ \\
       & $\frac{\rm{mJy}}{\rm{beam}^{-1}}$ & mJy         & M$_{\odot}$ & $\frac{10^{25}}{\rm{cm}^{-2}}$ & M$_{\odot}$ & $\frac{10^{25}}{\rm{cm}^{-2}}$ \\
       &                   &                & \multicolumn{2}{|c|}{H83$^a$}                & \multicolumn{2}{|c}{OH94$^a$} \\
\hline
7538IRS1mm1 & 1861 & 2990 & -- & -- & -- & -- \\
7538IRS1mm1$^{b,c}$ & 861 & 1990 & 115 & 18 & 43 & 7 \\
7538Smm1 & 24.1 & 132 & -- & -- & -- & -- \\
7538Smm1$^{b,d}$ & 16.1 & 124 & 38 & 1.8 & 14 & 0.7 \\
7538Smm1a  & 24.1 & 95  & -- & -- & -- & -- \\
7538Smm1a$^{b,d}$& 16.1 & 87  & 27 & 1.8 & 10 & 0.7 \\
7538Smm1b$^d$& 18.0 & 37  & 11 & 2.0 & 4  & 0.7 \\
7538Smm2$^d$ & 20.5 & 76  & 23 & 2.3 & 9  & 0.9 \\
7538Smm3$^d$ & 21.4 & 67  & 21 & 2.4 & 8  & 0.9 \\
\hline \hline
\end{tabular}
~\\
{\footnotesize $^a$ H83: \citet{hildebrand1983}; OH94: \citet{ossenkopf1994} \\ $^b$ Corrected for free-free flux contribution.\\ $^c$ Assumed temperature 245\,K.\\  $^d$ Assumed temperature 50\,K.\\ $^e$ Measured within the $4\sigma$ contours.}
\label{mm}
\end{table}

\begin{figure*}[htb] 
\includegraphics[width=18.4cm]{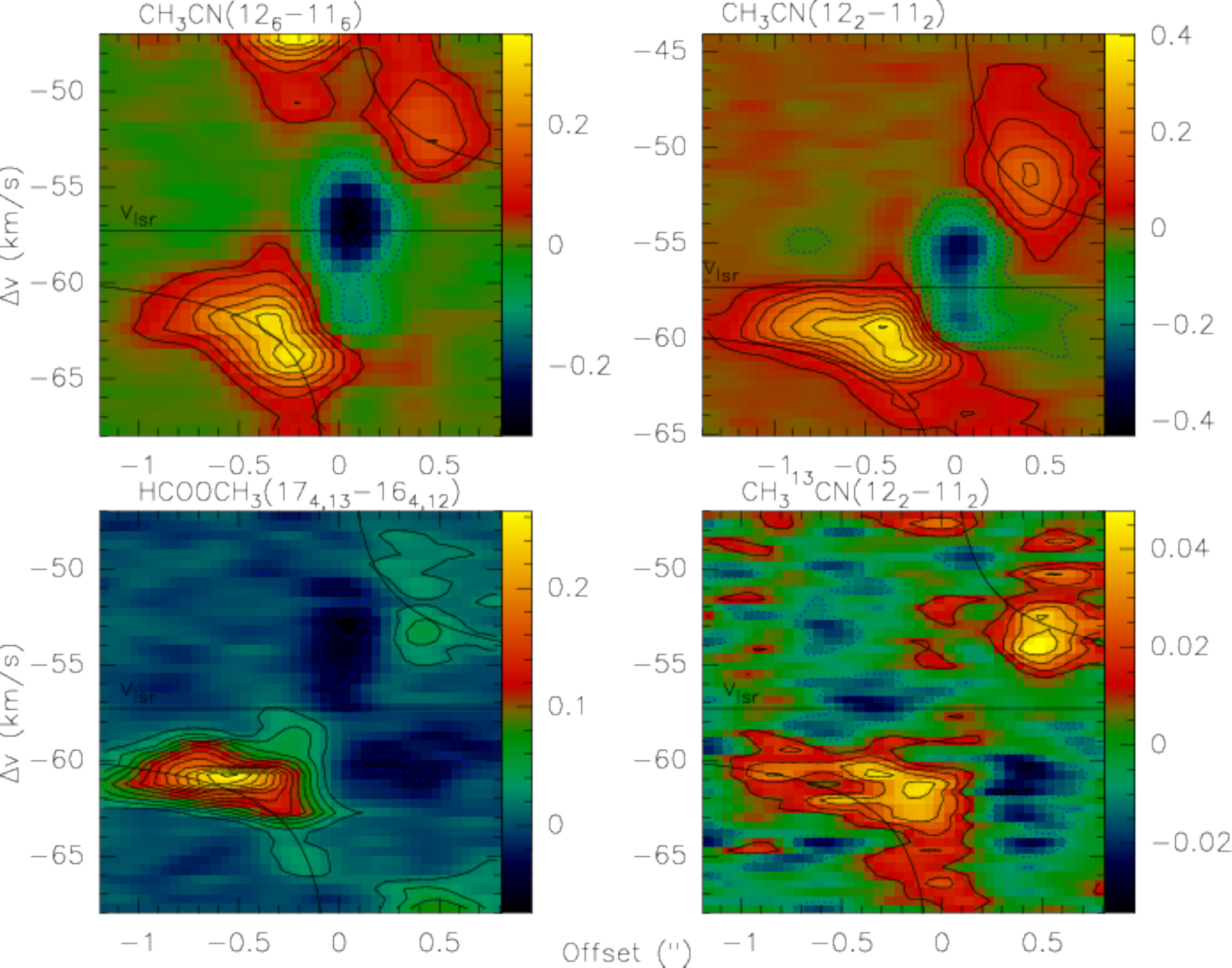}
\caption{Selected position-velocity diagrams for NGC7538IRS1 along the
  axis in Fig.~\ref{ch3cn_mom}. The presented molecules and their
  transitions are labeled above each panel, the approximate
  $v_{\rm{lsr}}\approx -57.3$\,km\,s${-1}$ is marked as well. The
  black lines in all panels correspond to Keplerian rotation curves
  around a 30\,M$_{\odot}$ central object. The units of the wedges are
  Jy beam$^{-1}$.}
\label{irs1_pv}
\end{figure*}

Assuming optically thin emission from dust following the standard
approach by \citet{hildebrand1983} we can estimate gas masses and
column densities at an assumed temperature. Because NGC7538IRS1 is a
strong infrared source and hot core, following \citet{qiu2011} we
assume a dust temperature of 245\,K for that source. NGC7538S is
supposedly younger and colder, and we assume a dust temperature of
50\,K for the corresponding sub-sources. Regarding the dust properties,
we calculate the masses and column densities following
\citet{hildebrand1983} (H83) on the one hand, and
\citet{ossenkopf1994} (OH94) for thin ice mantles at densities of
$10^6$\,cm$^{-3}$ on the other hand.  The gas-to-dust ratio is taken
as 186 following \citet{draine2007} and \citet{jenkins2004}.

Depending on the dust properties, toward NGC7538S we find core masses
between 4 and 38\,M$_{\odot}$ and column densities between $0.7\times
10^{25}$ and $1.8\times 10^{25}$\,cm$^{-2}$, corresponding to visual
extinctions on the order of $10^4$\,mag. While such extinctions are
very high, similar values have been reported in the past at
correspondingly high spatial resolution (e.g.,
\citealt{beuther2007d,rodon2008}). One should keep in mind that such
high extinction values are only found at the highest spatial
resolution achievable with interferometers. At lower resolution, the
emission smears out and lower values are found. Regarding the core
masses in NGC7538S, at first sight they do not appear extraordinarily
high, however, considering that approximately 90\% of the gas are
filtered out on larger scales, we only observe the densest structure
that is embedded in a much larger gas reservoir.

The situation is considerably different for NGC7538IRS1 where excessively
high column densities on the order of $10^{26}$\,cm$^{-2}$ are found
(corresponding to visual extinctions above $10^5$\,mag), as well as
core masses between 43 and 115\,$M_{\odot}$ (depending on the dust
properties) within a projected size of $\sim$2000\,AU. To the authors'
knowledge this is an extraordinary concentration of mass within small
spatial scales and will be discussed in more detail in section
\ref{cont_discuss_irs1}.

\begin{figure*}[htb] 
\includegraphics[width=18.4cm]{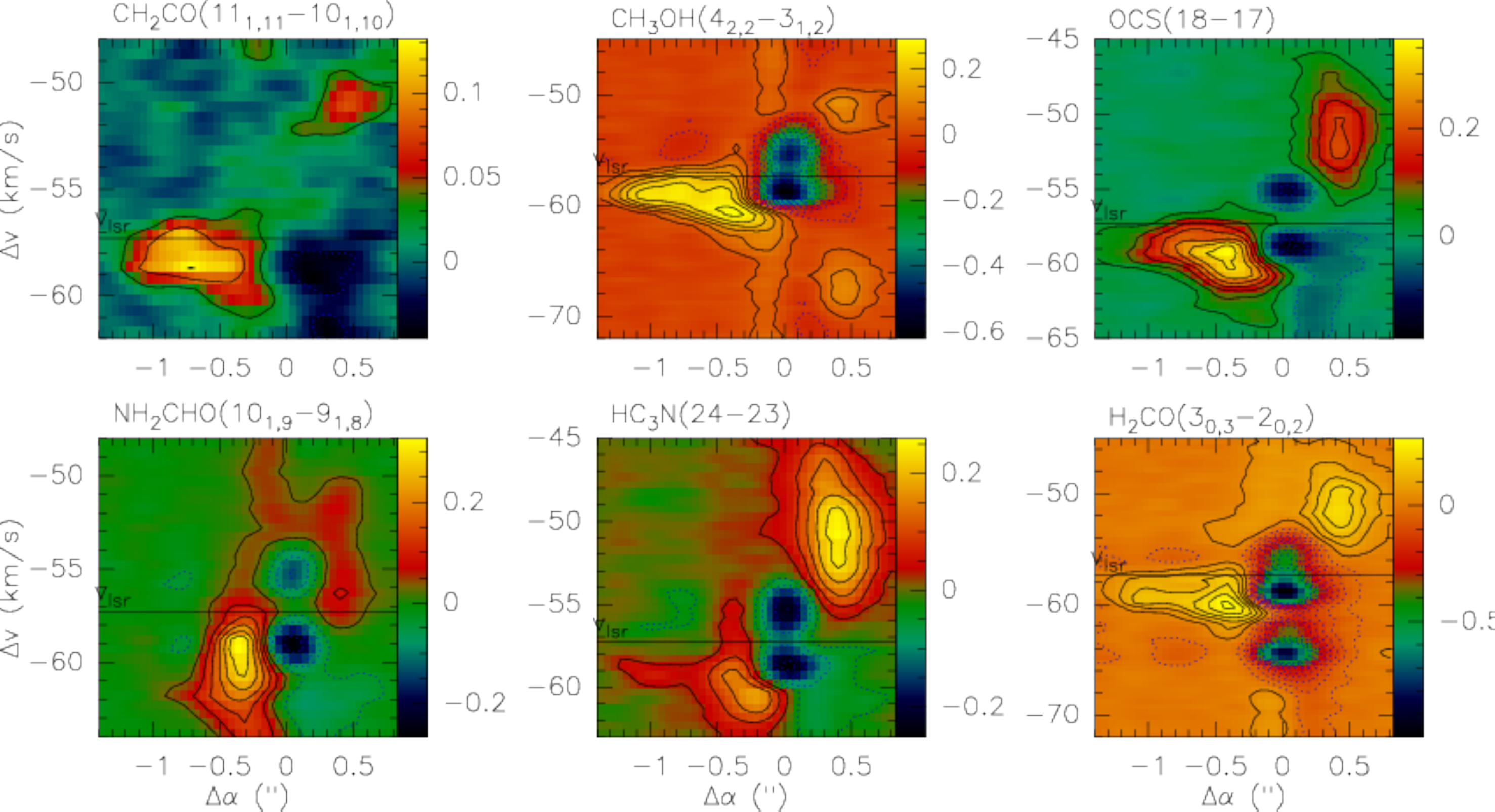}
\caption{Selected position-velocity diagrams for NGC7538IRS1 along the
  axis in Fig.~\ref{ch3cn_mom}. The presented molecules and their
  transitions are labeled above each panel, the approximate
  $v_{\rm{lsr}}\approx -57.3$\,km\,s$^{-1}$ is marked as well. The
  units of the wedges are Jy beam$^{-1}$.}
\label{irs1_pv2}
\end{figure*}

For comparison, we can also calculate the total gas masses of the two
regions based on the single-dish data. As approximate clump sizes, we
integrate the flux in the area within the 750\,mJy\,beam contour in
Figure \ref{overview}. For NGC7538IRS1 and NGC7538S, we get integrated
1.2\,mm fluxes of 20.9 and 6.2\,Jy, respectively. Following the same
approach as above, we can calculate the gas masses for the two dust
models H83 and OH94. On these large scale we use the temperature
estimates from \citet{sandell2004} who estimate 75 and 35 for
NGC7538IRS1 and NGC7538S. With these numbers we get total gas masses
for NGC7538IRS1 and NGC7538S of 2512 and 1757\,M$_{\odot}$ (H83) or
1011 and 706\,M$_{\odot}$ (OH94), respectively.

\subsection{Spectral line emission}
\label{lines}

All spectral lines listed in Table \ref{linelist} were detected toward
NGC7538IRS1 and most of them also toward NGC7538S. While also
velocity gradients are identified in both regions, the detailed
spectral line signatures between NGC7538IRS1 and NGC7538S are
considerably different, in particular we detect strong absorption
signatures toward NGC7538IRS1 but not toward NGC7538S.

\subsubsection{NGC7538IRS1}
\label{lines_ngc7538irs1}

Figure \ref{ch3cn_spectra_irs1} presents the CH$_3$CN$(12_k-11_k)$
($0\leq k \leq 3$) spectra toward the mm continuum peak as well as
toward a position approximately $1.1''$ offset south of the continuum
peak. While the offset spectrum is a typical CH$_3$CN emission
spectrum, the spectrum toward the continuum peak is dominated by
absorption features. While absorption features in interferometric data
should always be taken with caution because missing flux problems can
also artificially produce such features, the fact that we see the
absorption only toward the continuum peak but not toward an offset of
only $\sim 1.1''$ is a strong indicator for the absorption being a
real feature. While \citet{qiu2011} reported only redshifted
absorption in dense gas tracers at a spatial resolution of $\sim 3''
\times 2''$, and \citet{keto1991} and \citet{zheng2001} found only
blue-shifted absorption in lower-density NH$_3$ lines, Figure
\ref{ch3cn_spectra_irs1} already shows that at the high-spatial
resolution of our observations, the dense gas shows blue- and
red-shifted components simultaneously.

To identify potential rotational symmetries in these data, Figure
\ref{ch3cn_mom} presents the 1st and 2nd moment maps
(intensity-weighted peak velocities and line-widths) of our highest
excited CH$_3$CN$(12_6-11_6)$ line ($E_u/k=326$\,K). Although these
moment maps are affected by the absorption close to the continuum
peak, both maps clearly identify a velocity gradient in
northeast-southwest direction, consistent with the previous
lower-resolution Submillimeter Array observations in OCS and SO$_2$ by
\citet{klaassen2009}. While the 1st moment map exhibits a blue-red
velocity gradient extending about 10\,km\,s$^{-1}$ which is
perpendicular to the northwest-southeast outflow structure reported by
\citet{davis1998} and \citet{qiu2011}, it is interesting that also the
line-width map shows a significant line-width increase close to
approximately this axis while the line-widths are considerably smaller
northwest and southeast of that.  The 2nd moment map in Figure
\ref{ch3cn_mom} gives visually the impression of a disk-like
structure, however, again this needs to be taken with caution because
that signature can be affected by the absorption of the gas against
the strong continuum.

\begin{figure*}[t!] 
\includegraphics[width=9cm]{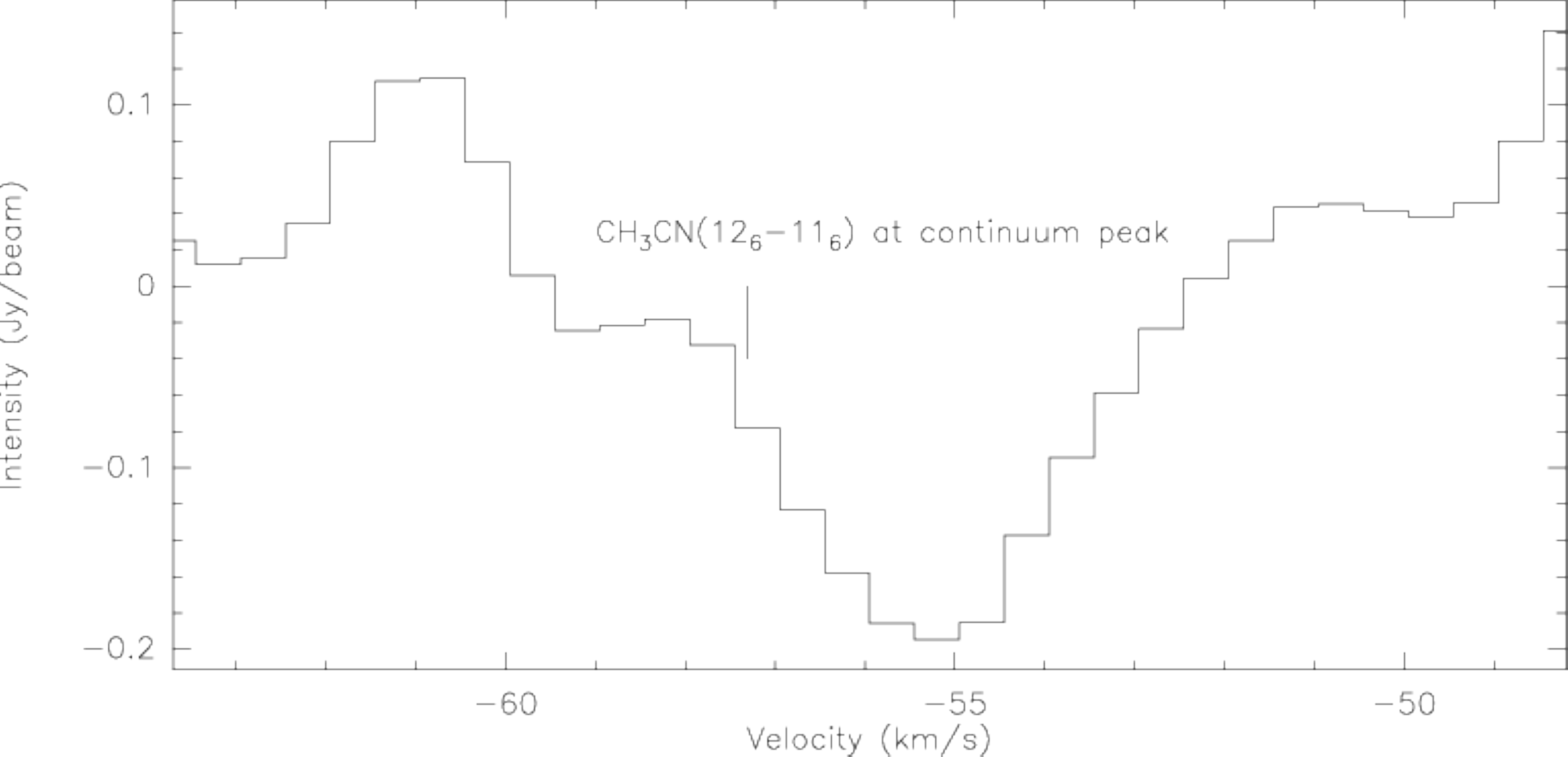}
\includegraphics[width=9cm]{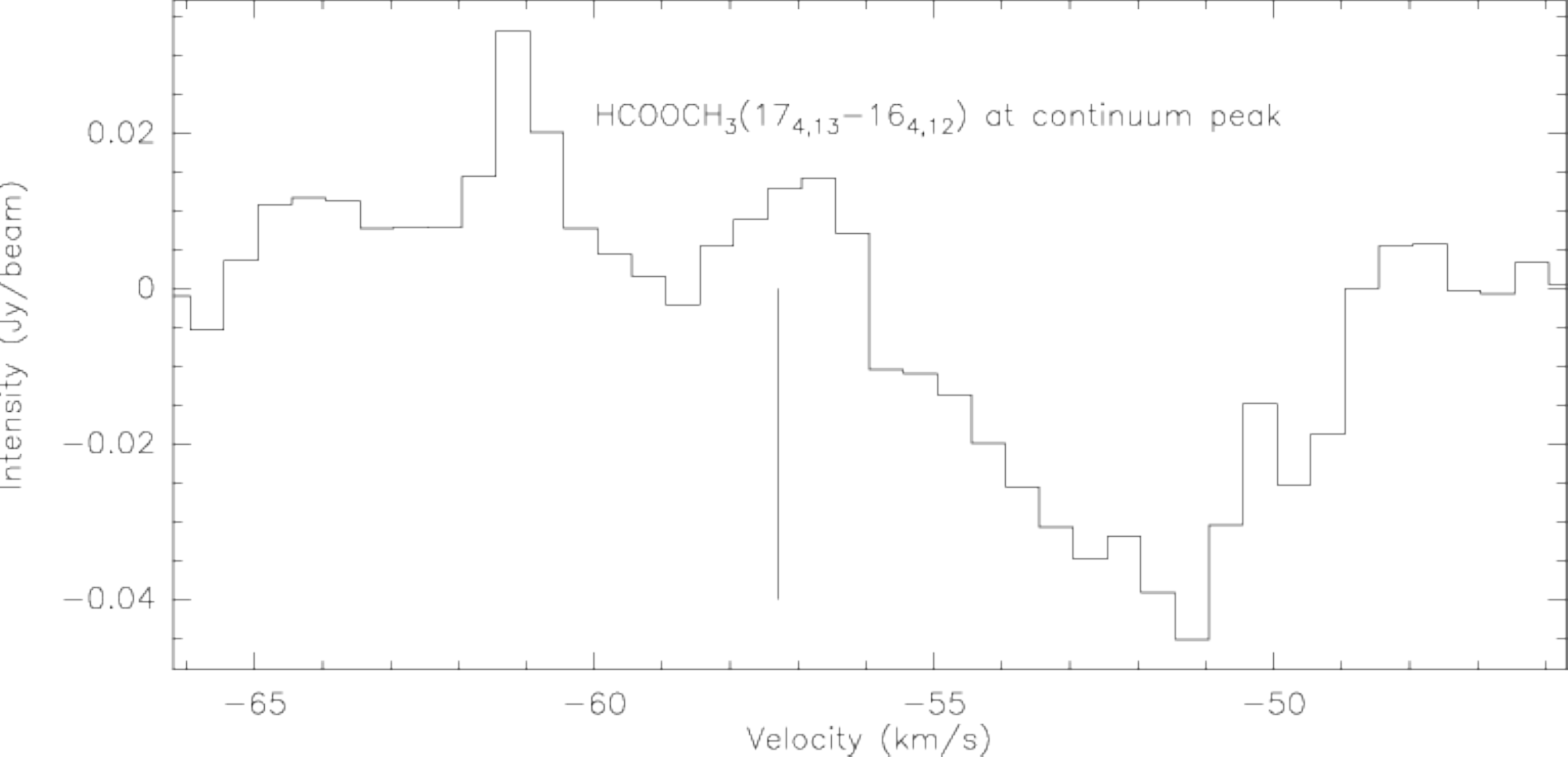}
\includegraphics[width=9cm]{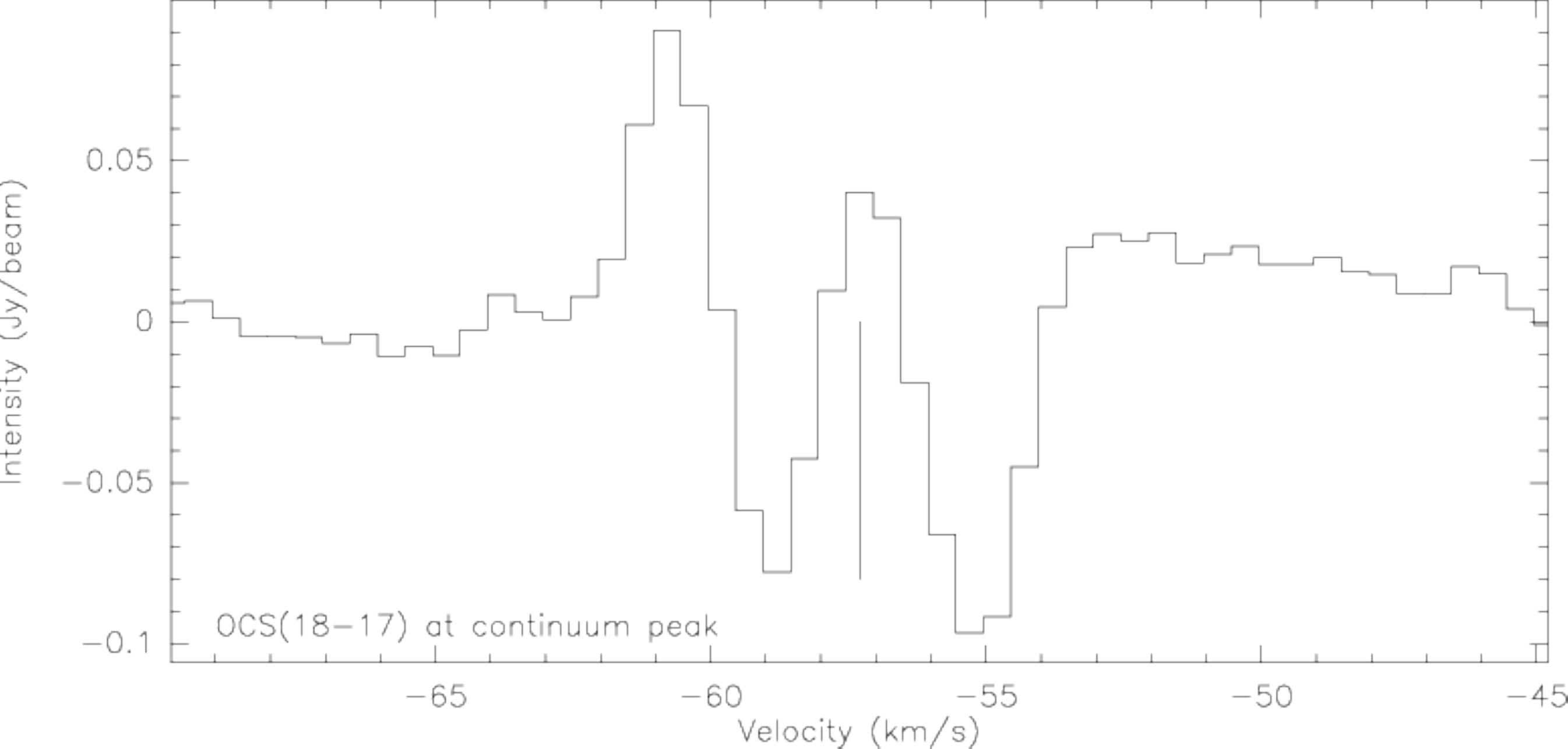}
\includegraphics[width=9cm]{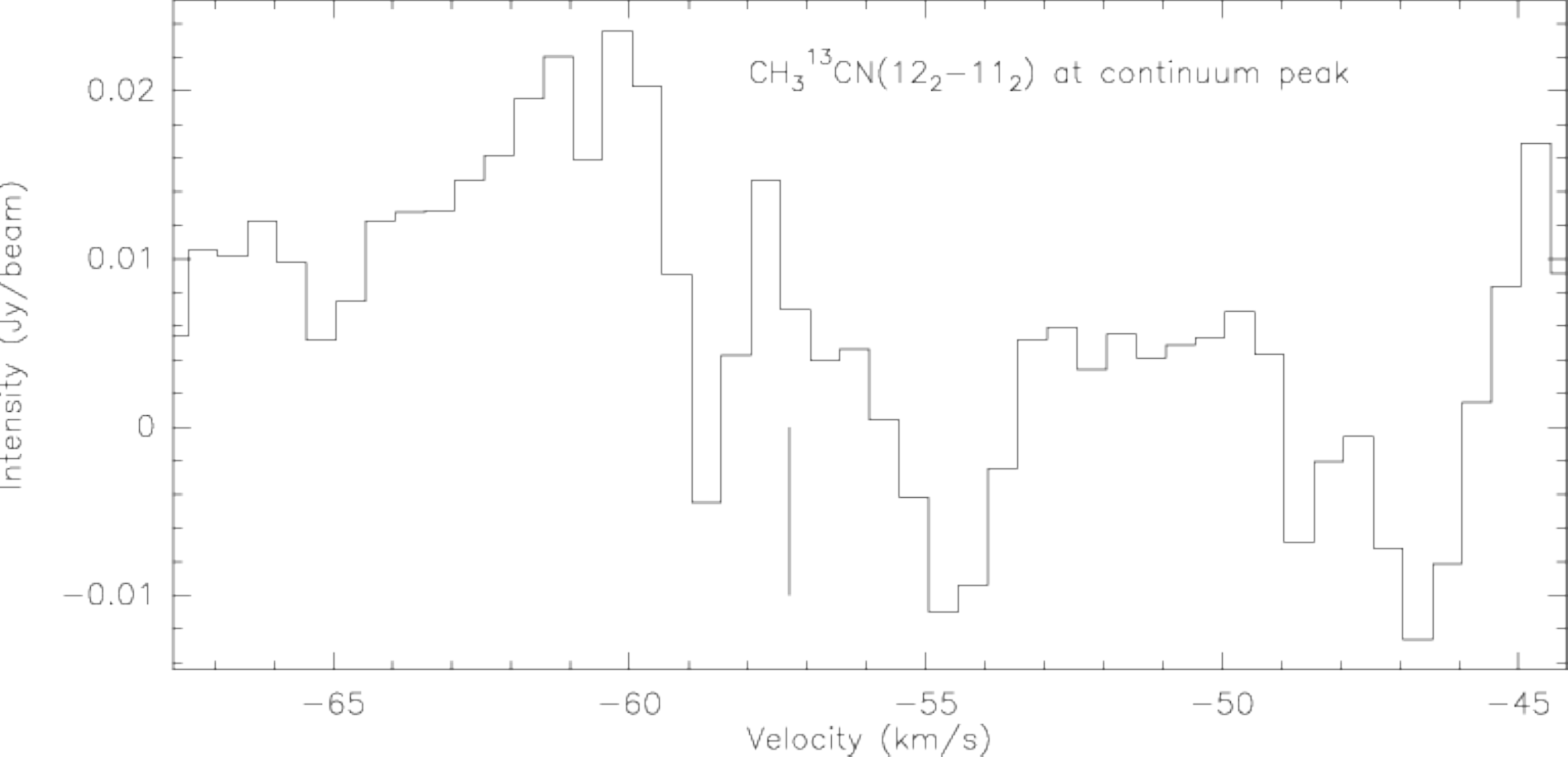}
\includegraphics[width=9cm]{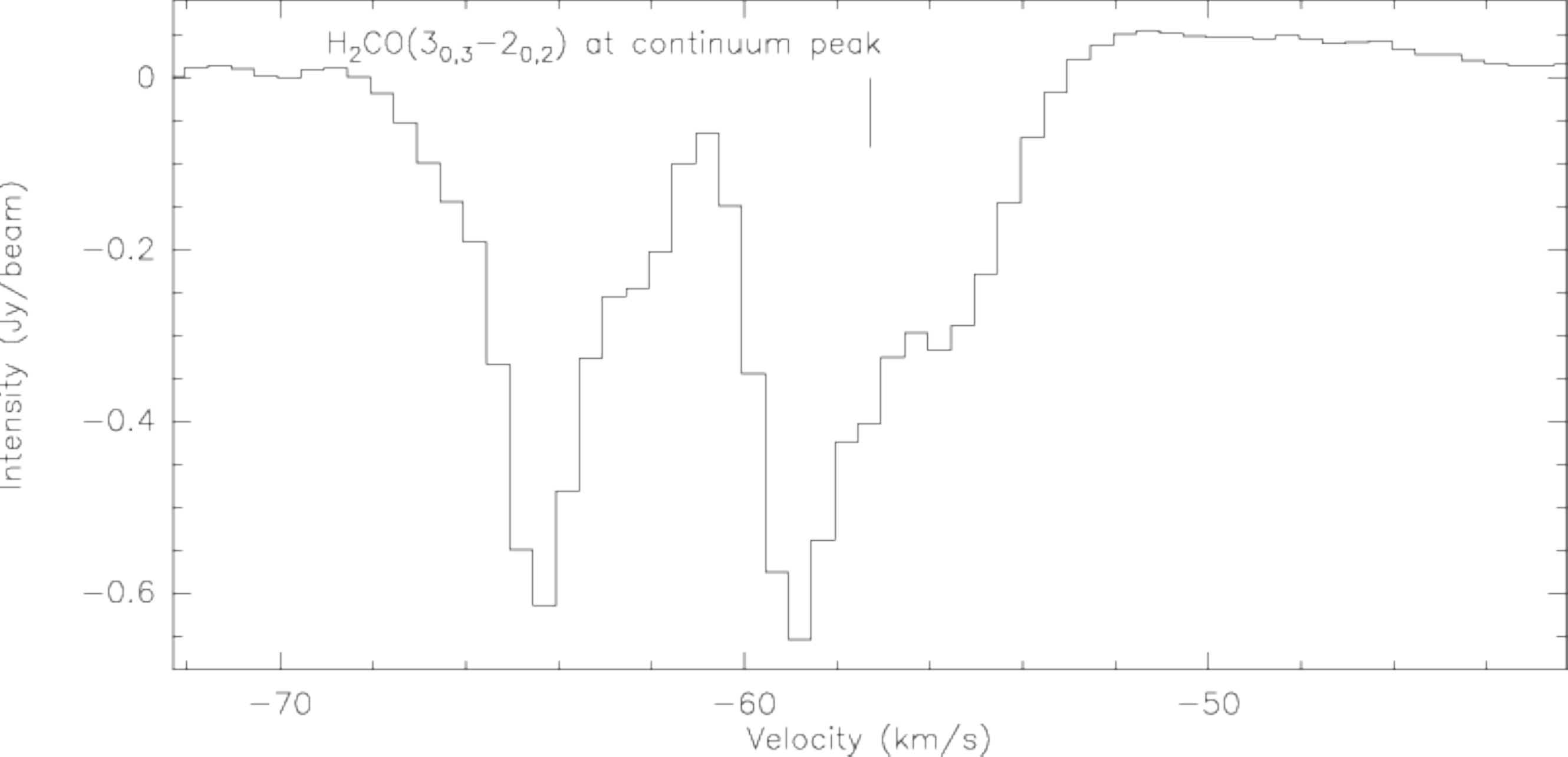}
\includegraphics[width=9cm]{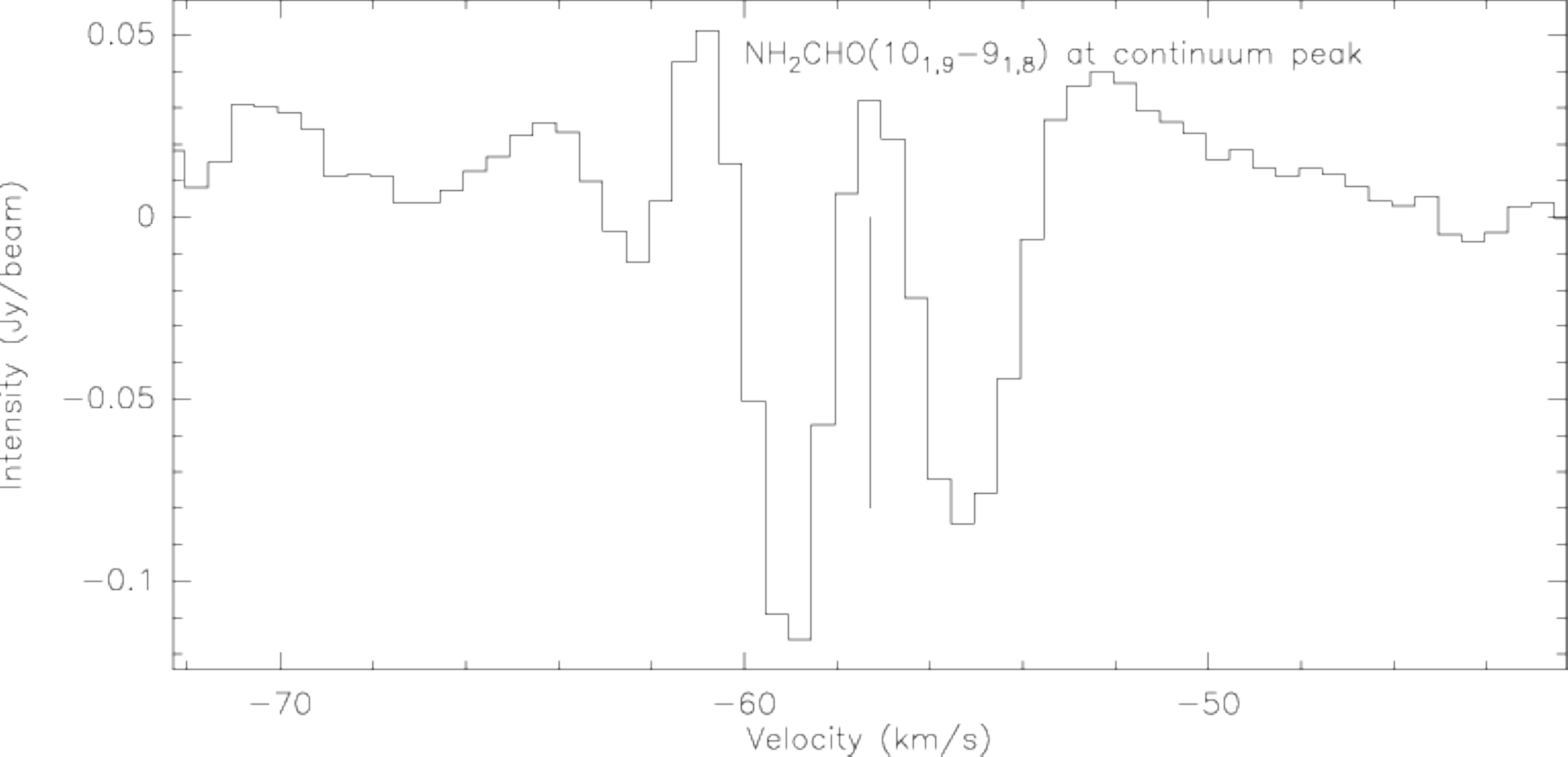}
\includegraphics[width=9cm]{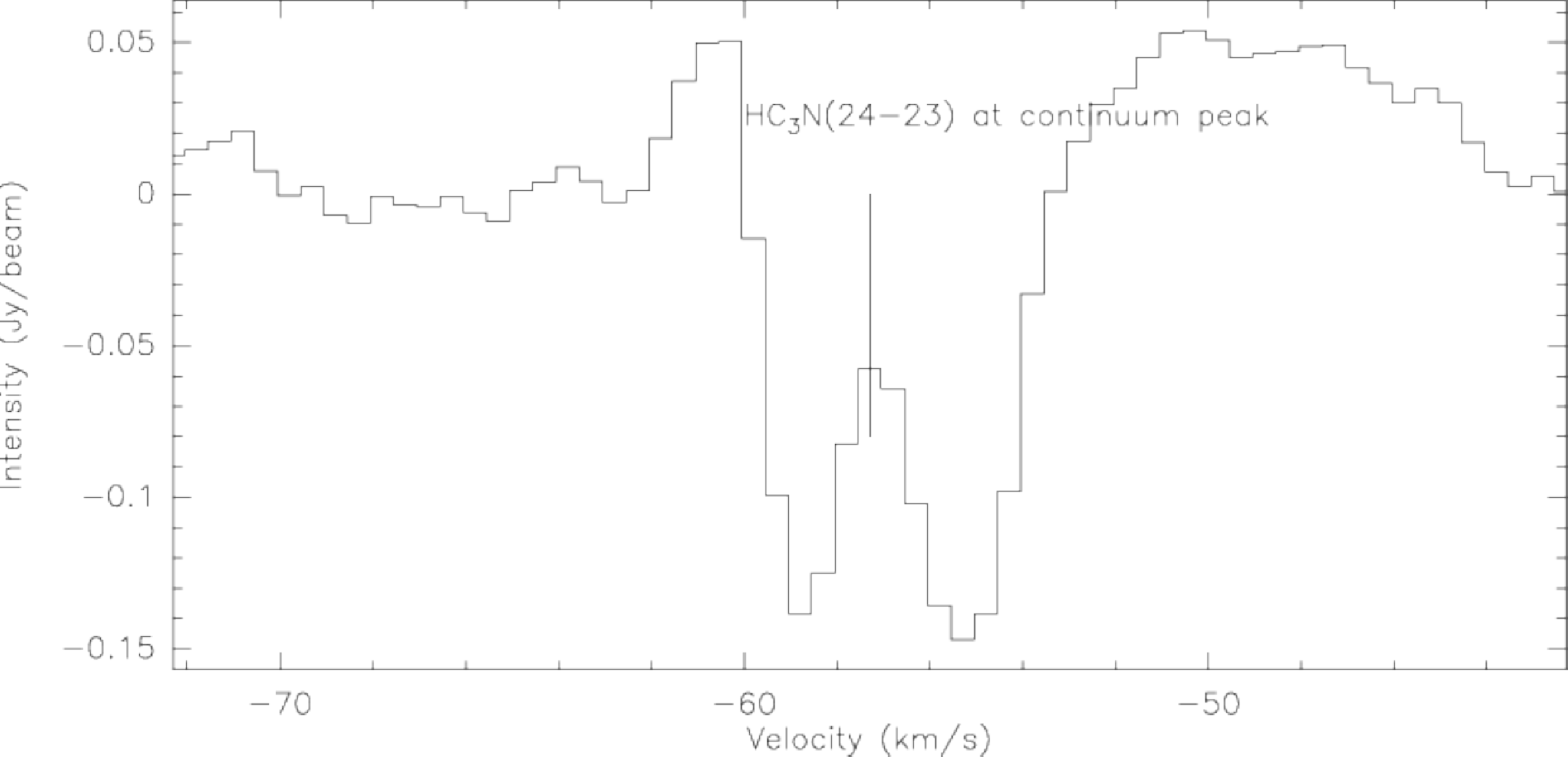}
\includegraphics[width=9cm]{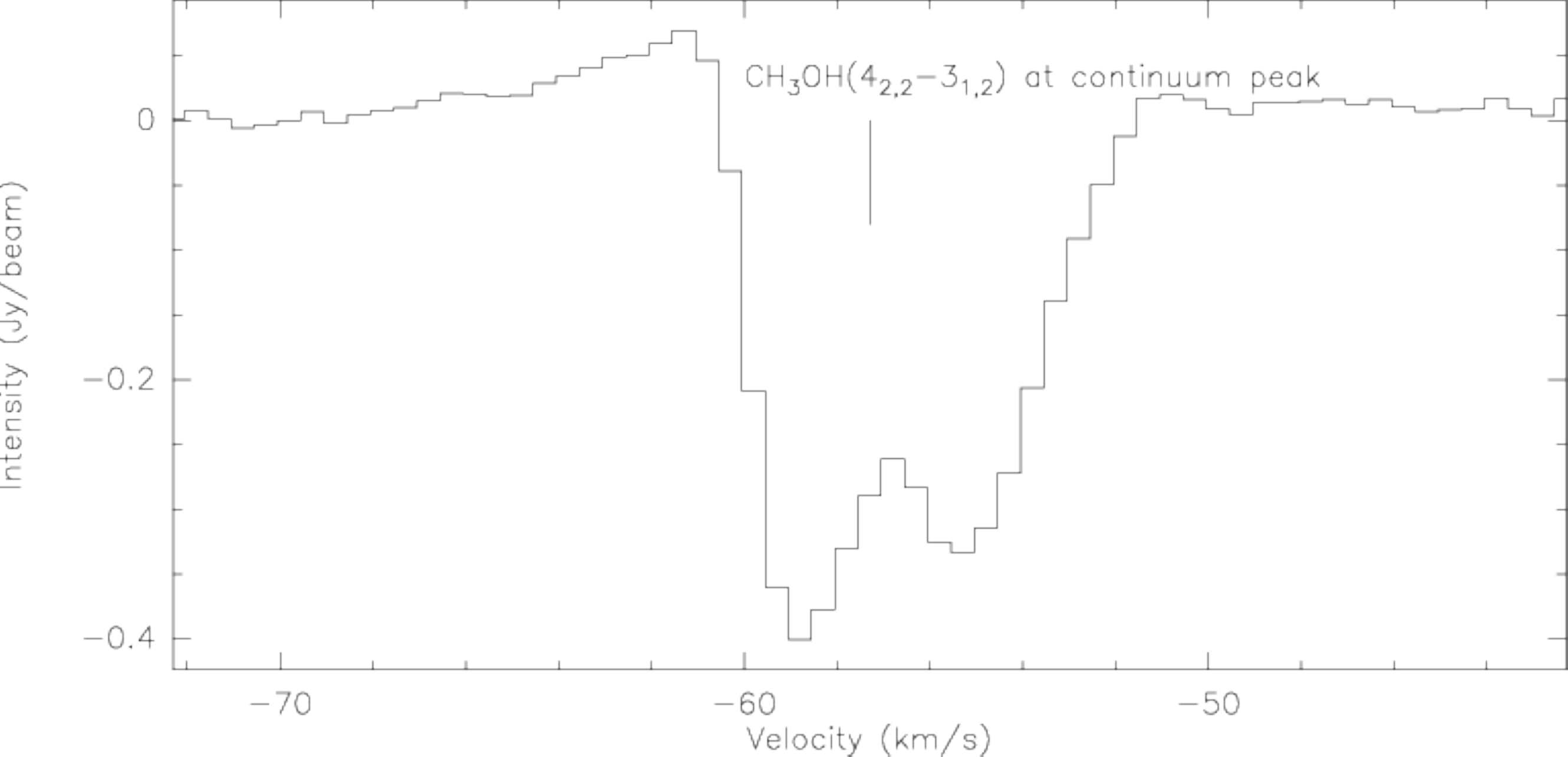}
\includegraphics[width=9cm]{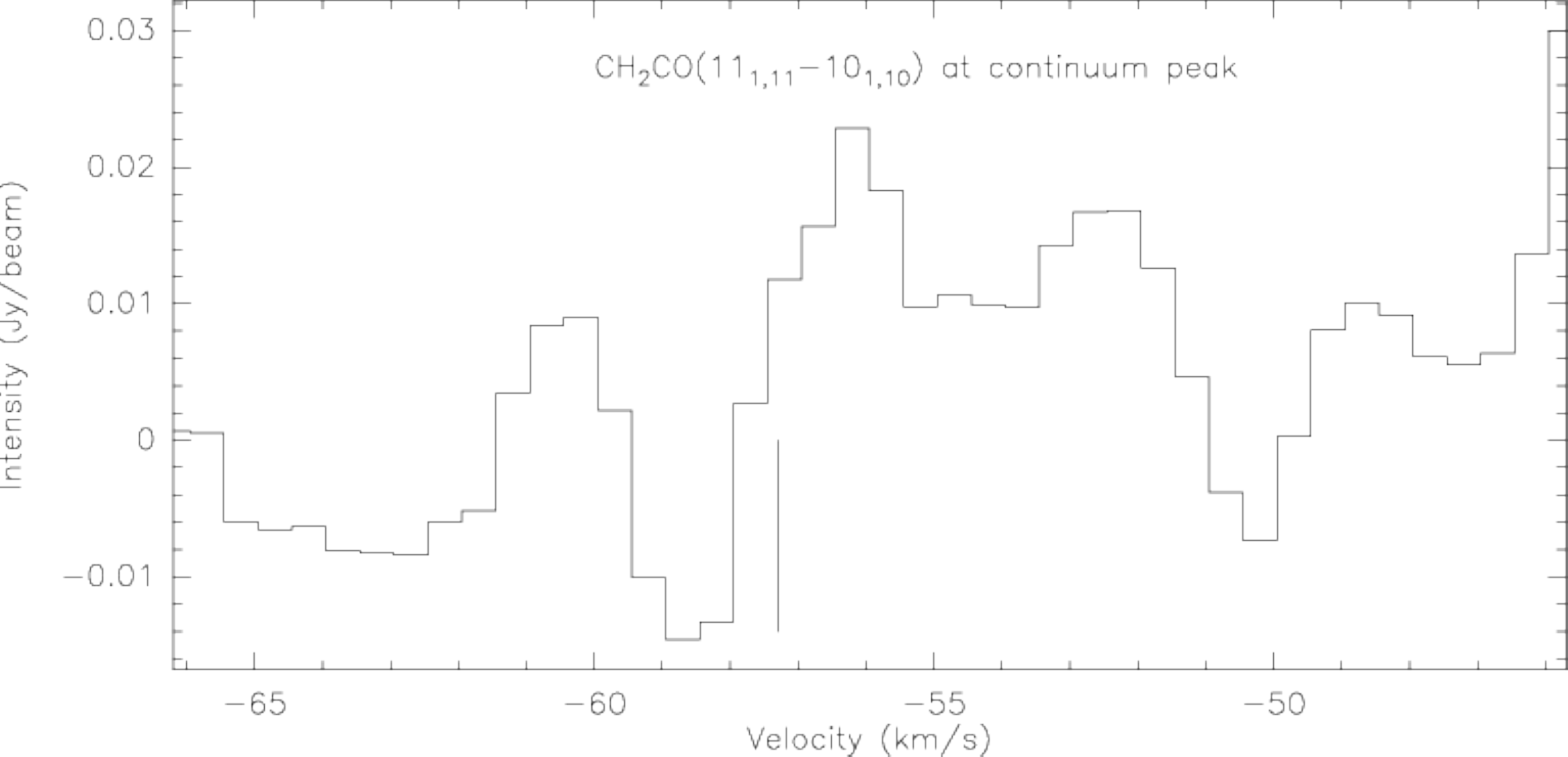}
\caption{Spectra toward the 1.36\,mm continuum peak of NGC7538IRS1 for
  selected lines as marked in each panel. The velocity of rest is
  marked by a line.}
\label{spectra_irs1_2}
\end{figure*}

\begin{figure*}[t!] 
\includegraphics[width=18.4cm]{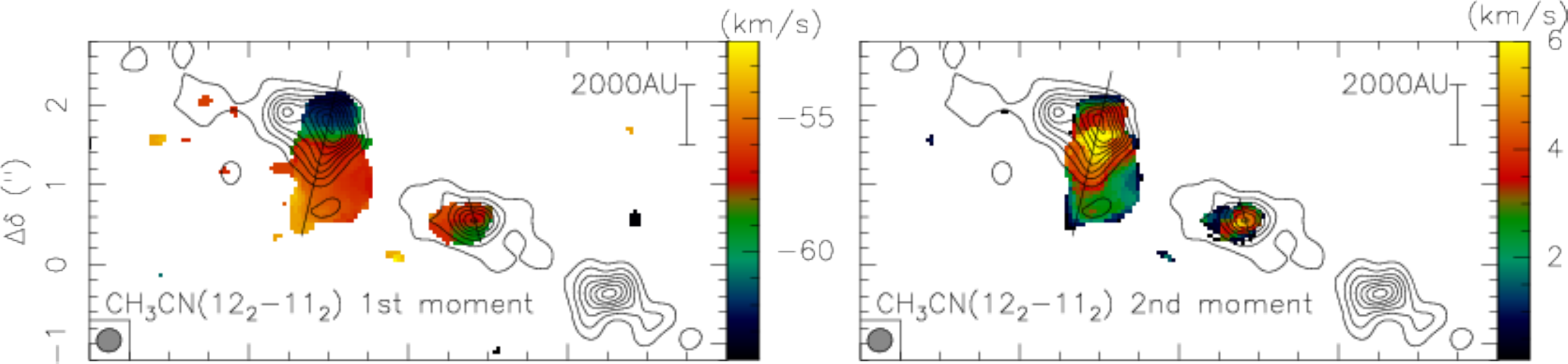}
\includegraphics[width=18.4cm]{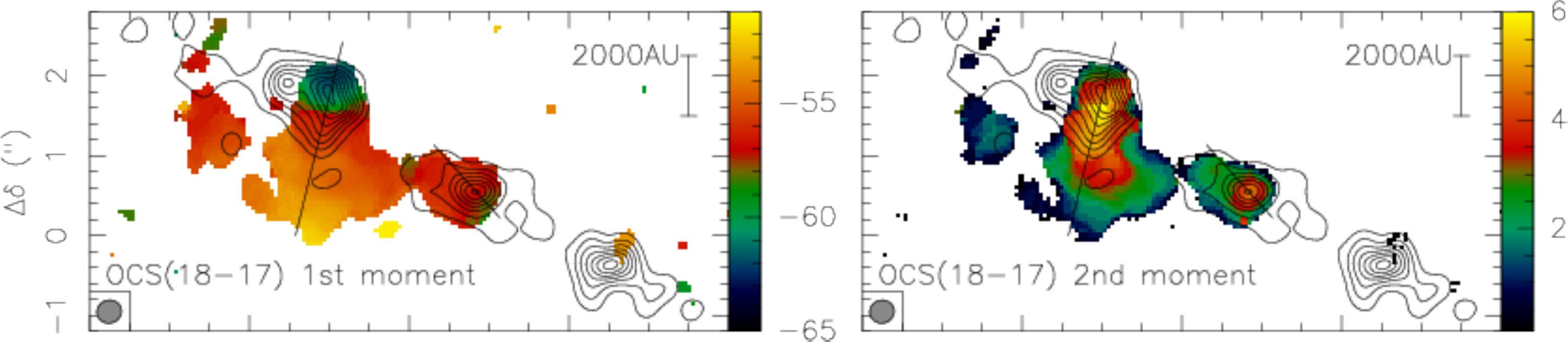}
\includegraphics[width=18.4cm]{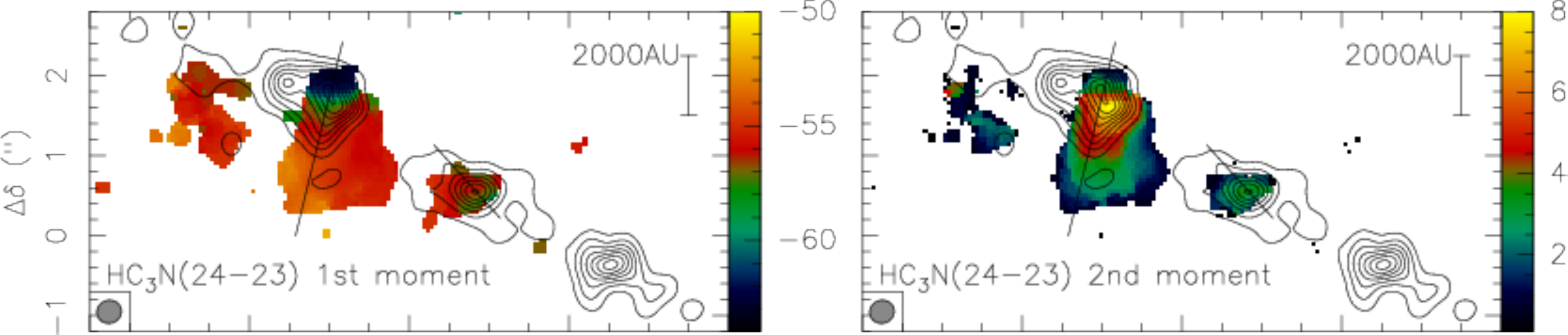}
\includegraphics[width=18.4cm]{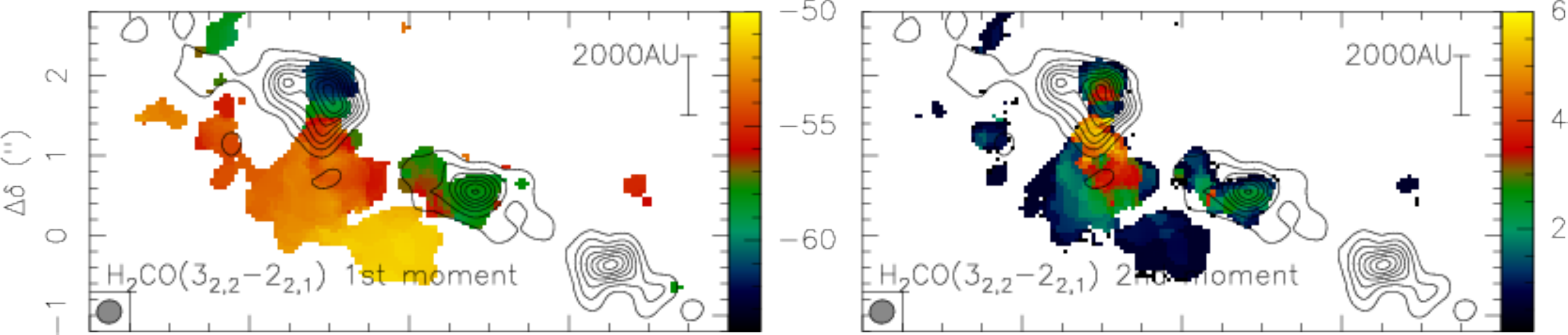}
\includegraphics[width=18.4cm]{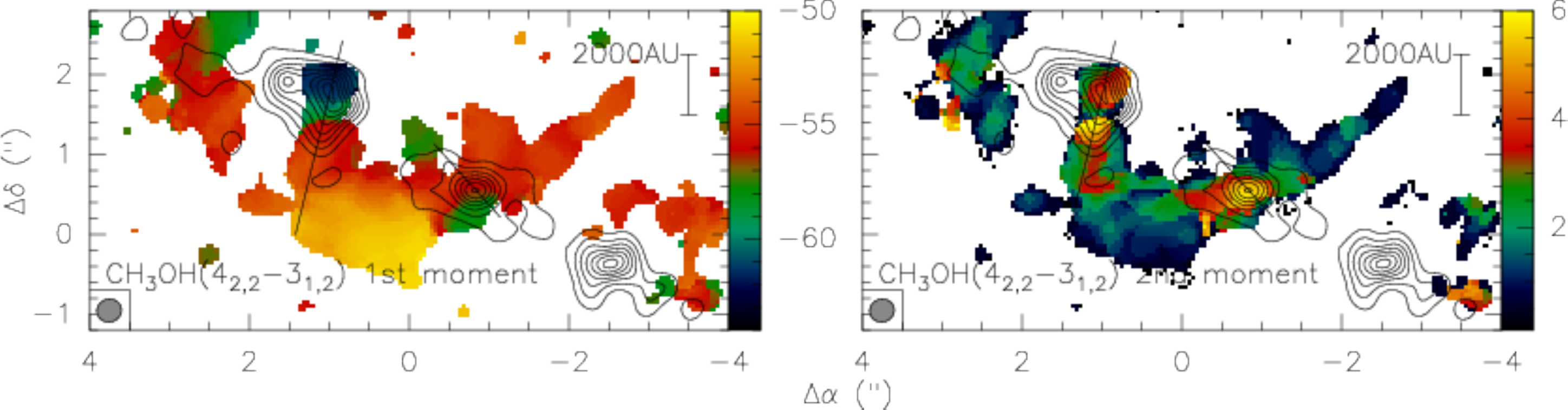}
\caption{The color scale presents in the left and right panels the
  first and second moment maps (intensity weighted peak velocities and
  line widths) toward NGC7538S. The molecular lines are marked in each
  panel. The contours show the 1.36\,mm continuum in $4\sigma$ steps
  ($1\sigma\sim 0.7$\,mJy\,beam$^{-1}$). The synthesized beam and a
  scale-bar are presented in each panel. The lines outline the axis
  used for the pv-cuts in Figures \ref{7538s_mm1_pv1} and
  \ref{7538s_mm2_pv}.}
\label{7538s_overlay}
\end{figure*}

Figures \ref{irs1_pv} and \ref{irs1_pv2} now show the
position-velocity diagrams along the northeast-southwest cut outlined
in Figure \ref{ch3cn_mom}. The typical hot molecular core and
high-density gas tracers CH$_3$CN and HCOOCH$_3$ exhibit absorption
signatures that are dominated by a red-shifted component but show some
blue-shifted absorption as well. This signature appears rather
independent of the excitation temperature because the shown $k=2$ and
$k=6$ CH$_3$CN$(12_k-11_k)$ components cover a range in excitation
temperatures $E_u/k$ of 250\,K (see Table \ref{linelist}). As
expected, the more optically thin isotopologues CH$_3^{13}$CN does not
exhibit such absorption features. While the red-shifted emission part
of these spectra is consistent with a typical Keplerian rotation
structure, the blue-shifted part of the emission spectrum does not
show such a signature. Figure \ref{irs1_pv} also presents a Keplerian
curve for a 30\,M$_{\odot}$ central object (see section \ref{intro}),
again showing the reasonable agreement on the red part of the spectrum
but not on the blue side. It is also interesting to note that the
higher excited CH$_3$CN$(12_6-11_6)$ shows on average a better
agreement with the Keplerian curve than the lower excited
CH$_3$CN$(12_2-11_2)$ line.  This is likely due to the fact that the
higher excited line traces gas closer to the star which hence exhibits
higher velocities. In contrast to that, one can also argue that the
non-correspondence of the blue part of the spectrum with Keplerian
rotation is not much of a surprise because Keplerian rotation implies
that the whole structure is dominated by the central object. This is
clearly not the case considering that the central star should have a
mass of $\sim$30\,M$_{\odot}$ (see Introduction) and the gas mass of the
central structure traced by the mm continuum emission is of that order
or even higher as well (Table \ref{mm}).

\begin{figure*}[htb] 
\includegraphics[width=9cm]{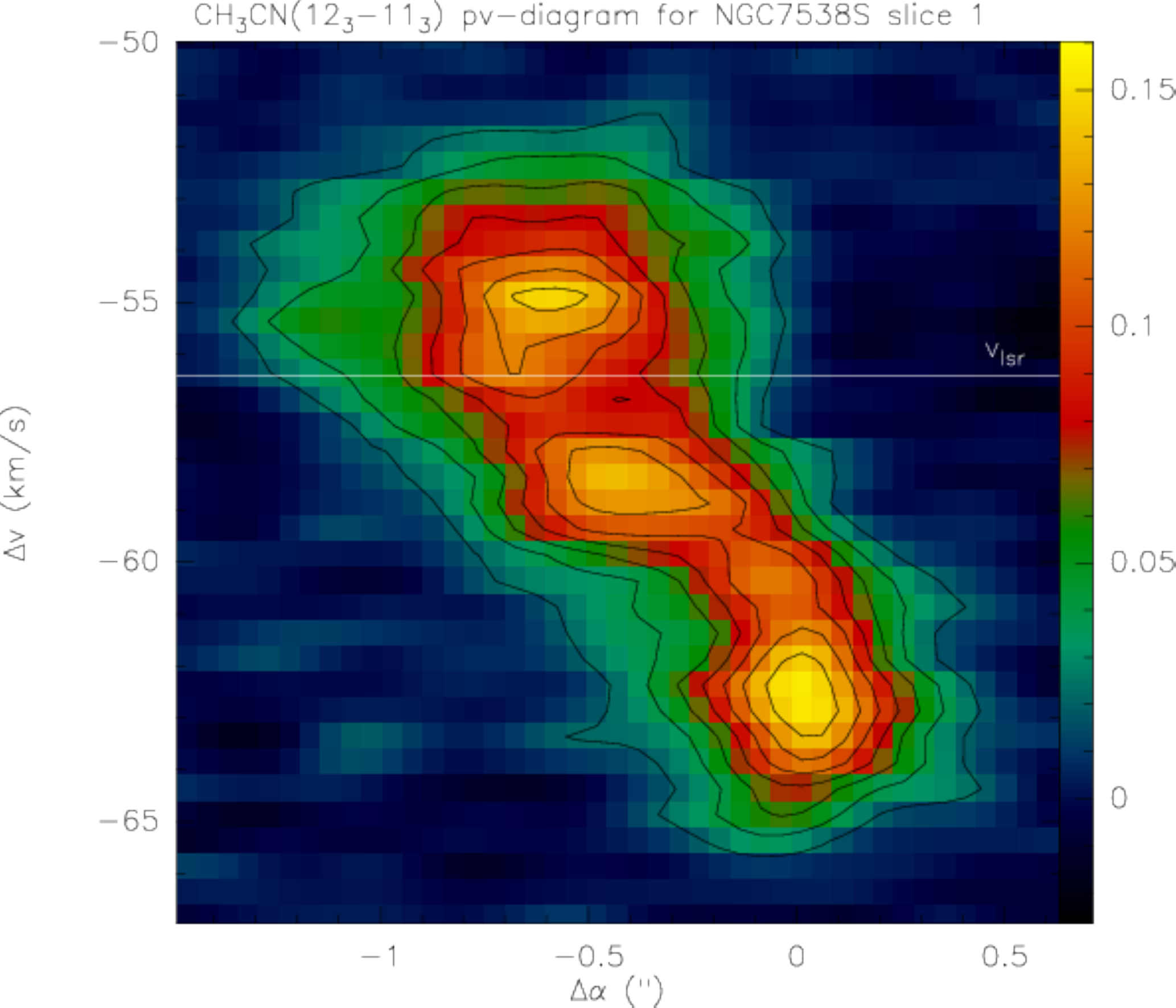}
\includegraphics[width=9cm]{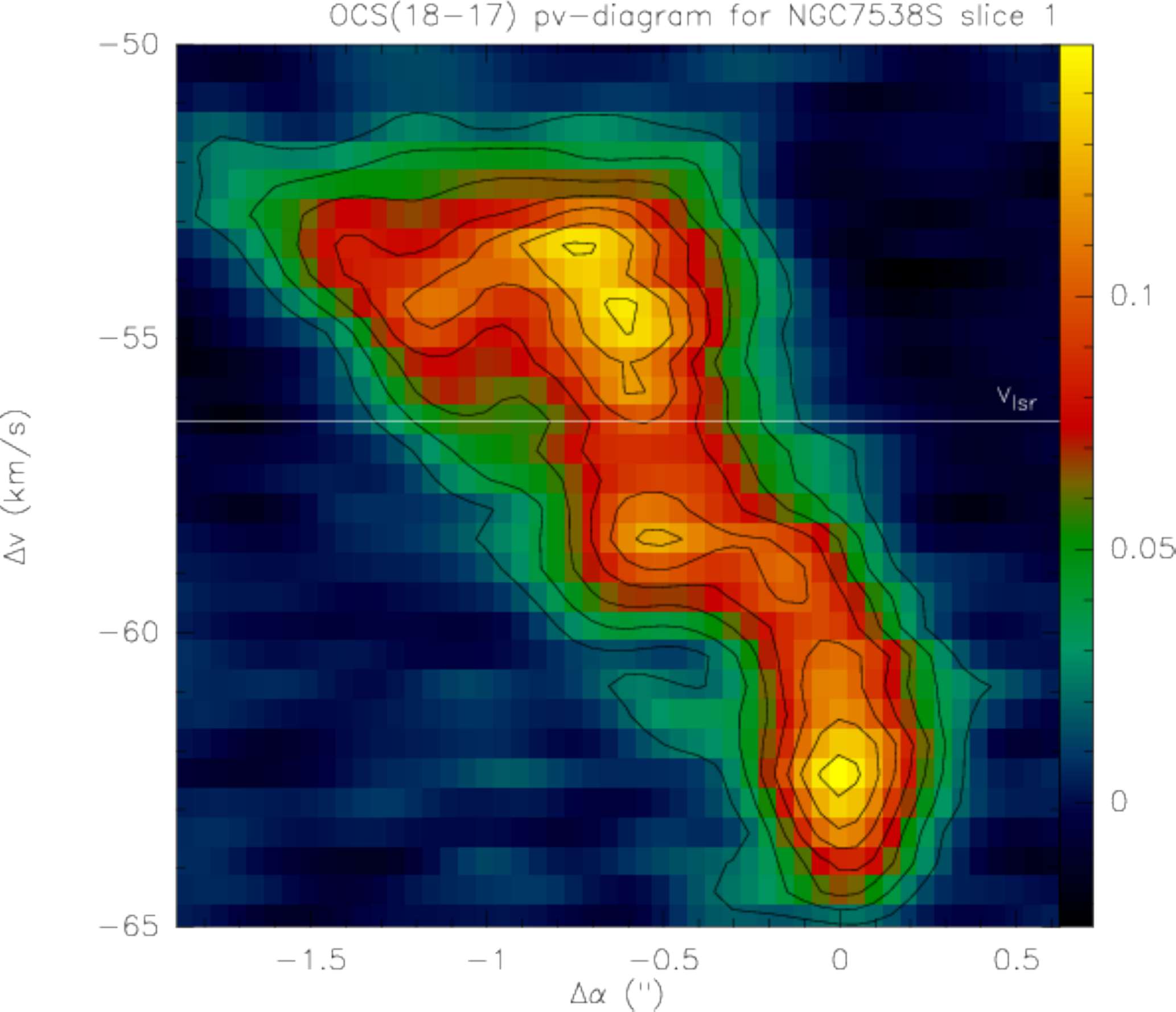}
\includegraphics[width=9cm]{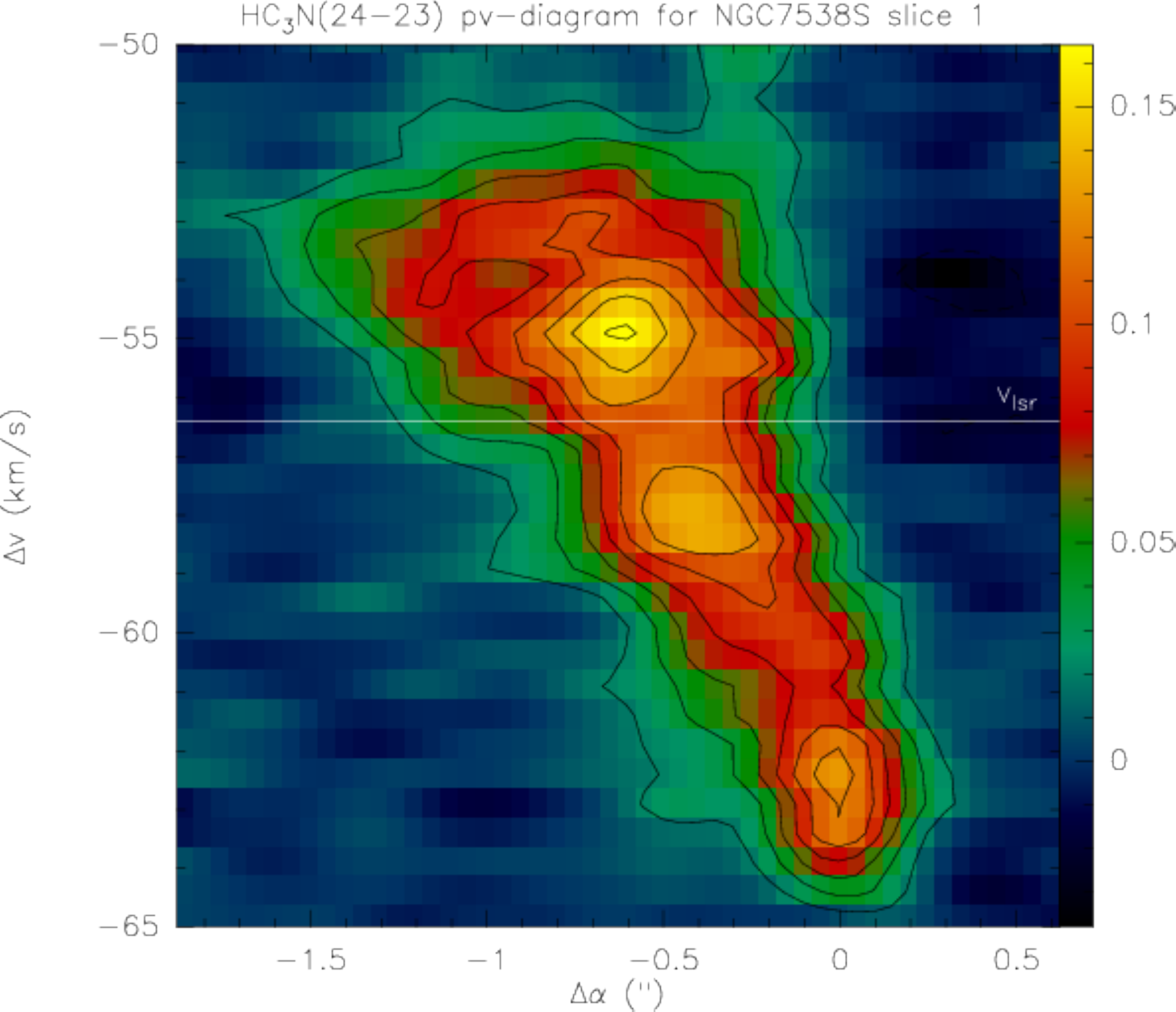}
\includegraphics[width=9cm]{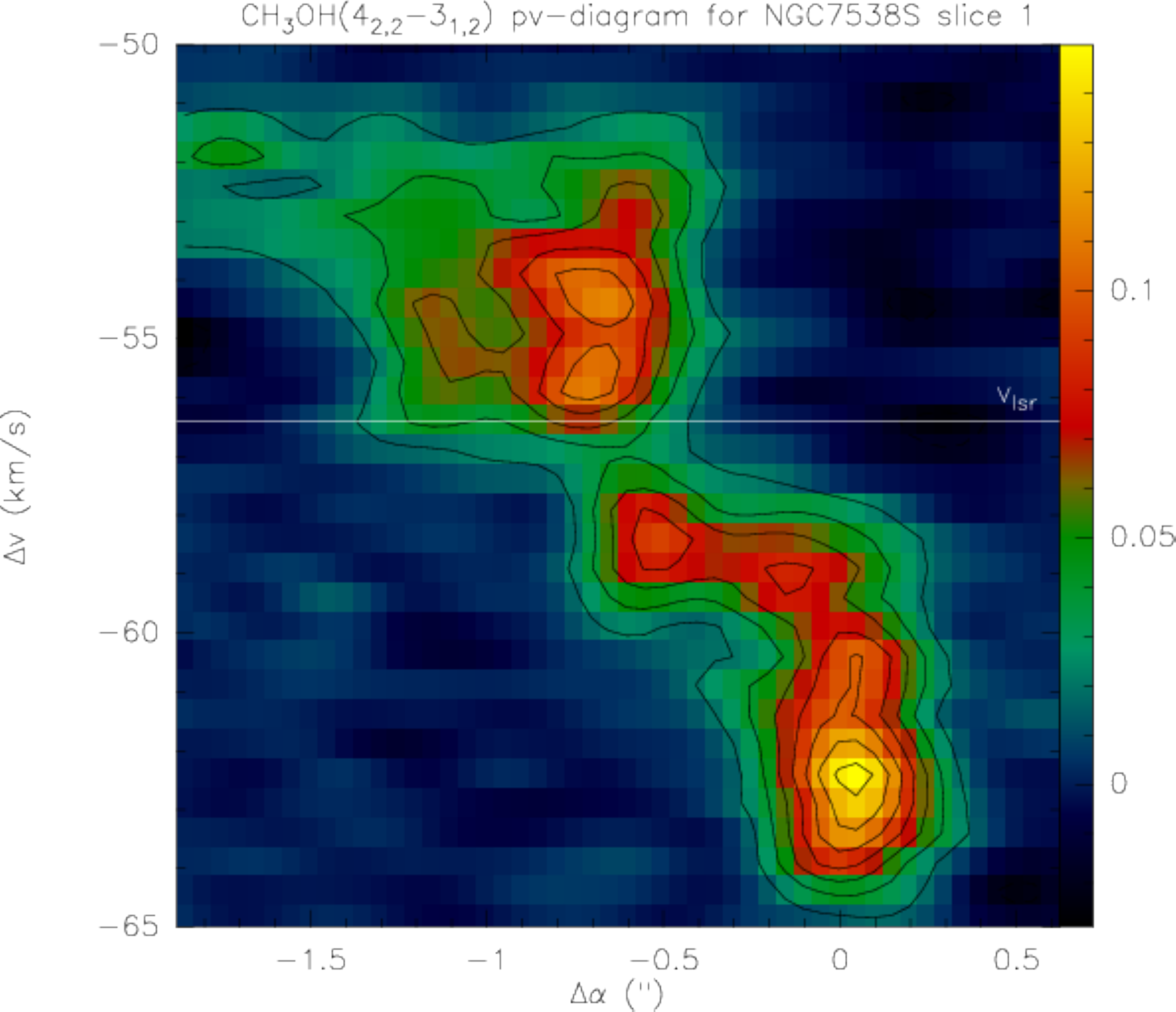}
\caption{Position velocity cuts through NGC7538S mm1 in several lines
  as marked in each panel along the axis shown in
  Fig.~\ref{7538s_overlay}, respectively. The $v_{\rm{lsr}}$ is
  marked. Offset 0 corresponds to the peak of mm1a. The units of the
  wedges are Jy beam$^{-1}$.}
\label{7538s_mm1_pv1}
\end{figure*}

The position-velocity diagrams presented in Figure \ref{irs1_pv2} show
slightly different absorption signatures. While almost all of them
show red-shifted absorption as well (except CH$_2$CO), the
blue-shifted absorption is at least as strong as that, if not
stronger. From a blue-shifted perspective, the lowest-excited H$_2$CO
line shows a particularly interesting feature because in addition to
the blue-shifted component at around -59\,km\,s$^{-1}$, it exhibits
another blue-shifted component at even more negative velocities around
-65\,km\,s$^{-1}$. A similar feature was recently also reported in the 1
and 2\,cm H$_2$CO lines (AAS poster by \citealt{yuan2011}).

A different way to investigate the various absorption features is
presented in Figure \ref{spectra_irs1_2} where we show the spectra of
the various lines extracted directly toward the mm continuum peak
position. The absorption features discussed in the paragraphs above
are exactly recovered there. Many of the spectra clearly show a
double-dibbed signature red- and blue-shifted around the
$v_{\rm{lsr}}$. To check whether the additional higher-velocity
absorption component in the H$_2$CO line is real absorption against
the continuum or rather due to missing flux on larger scales,
similarly as shown for CH$_3$CN in Figure \ref{ch3cn_spectra_irs1}, we
also extracted the H$_2$CO spectrum toward the position $0.2''/-1.1''$
to the south. And like for CH$_3$CN, the H$_2$CO spectrum exhibits a
pure and ``normal'' emission spectrum at that position. This implies
that the additional absorption at $\sim -65$\,km\,s$^{-1}$ should be
real.  Implications of the observed red- and blue-shifted absorption
toward NGC7538IRS1 will be discussed in section
\ref{kinematics_ngc7538irs1}.

\subsubsection{NGC7538S}
\label{lines_ngc7538s}

Toward the second region NGC7538S we clearly detect all CH$_3$CN
lines, as well as the spectral lines from OCS, HC$_3$N, H$_2$CO and
CH$_3$OH. In contrast to that, HCOOCH$_3$, NH$_2$CHO and CH$_2$CO are
barely detected. There is only a tentative detection of the latter two
molecules toward mm2. Regarding the clearly detected molecules and
spectral lines, it is interesting that all of them are detected toward
the two mm sub-peaks mm1 and mm2 but none of them toward the third mm
peak mm3. This already indicates peculiar chemical and evolutionary
differences between mm1 and mm2 on the one side and mm3 on the other
side. Furthermore, within mm1 we always detect mm1a in the spectral
line emission but no molecular line is found toward mm1b. A detailed
spectral and chemical analysis of all the other broadband line data we
observed simultaneously will be presented in a forthcoming paper. Here
we concentrate on the kinematics of the mm peaks mm1 and mm2.

\begin{figure*}[htb] 
\includegraphics[width=9cm]{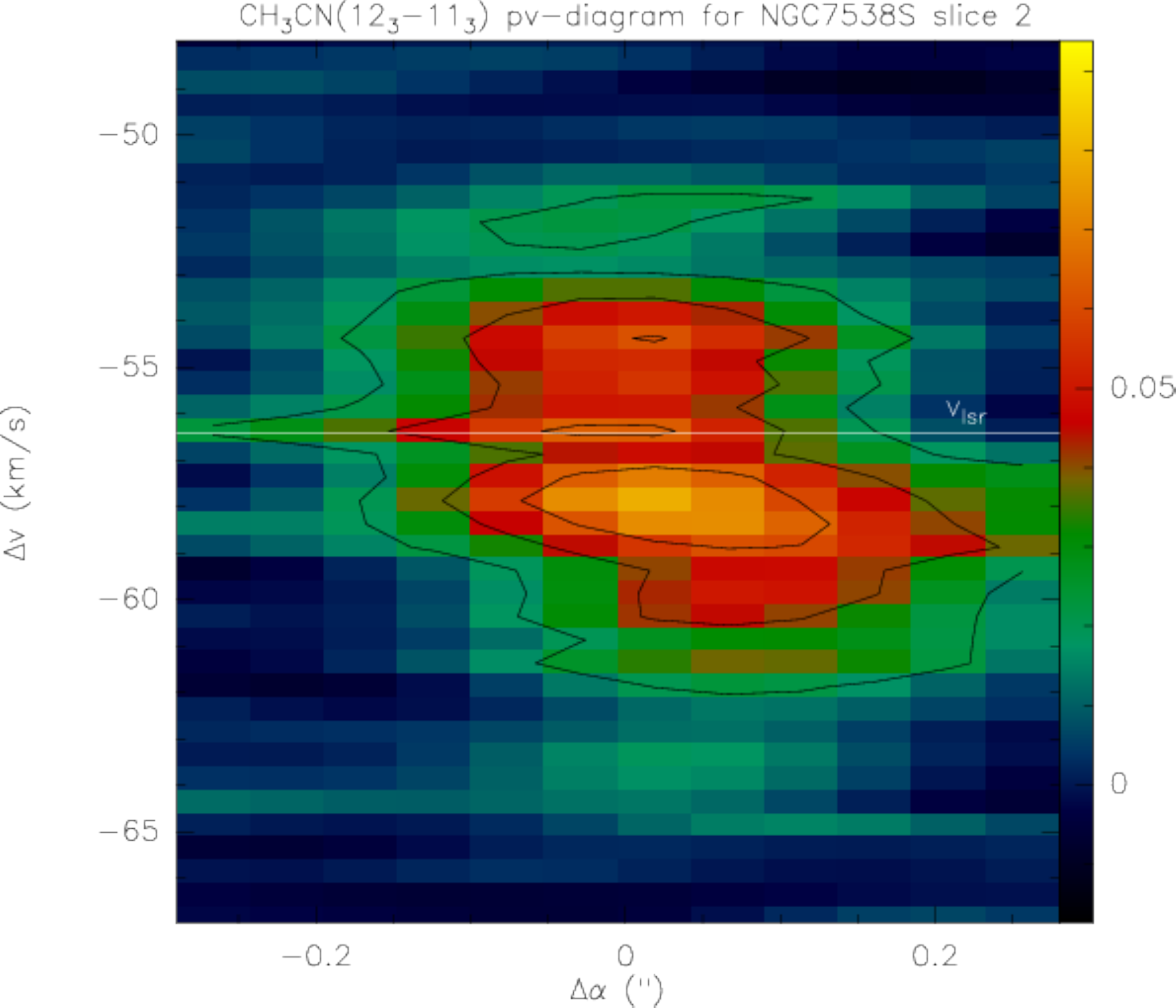}
\includegraphics[width=9cm]{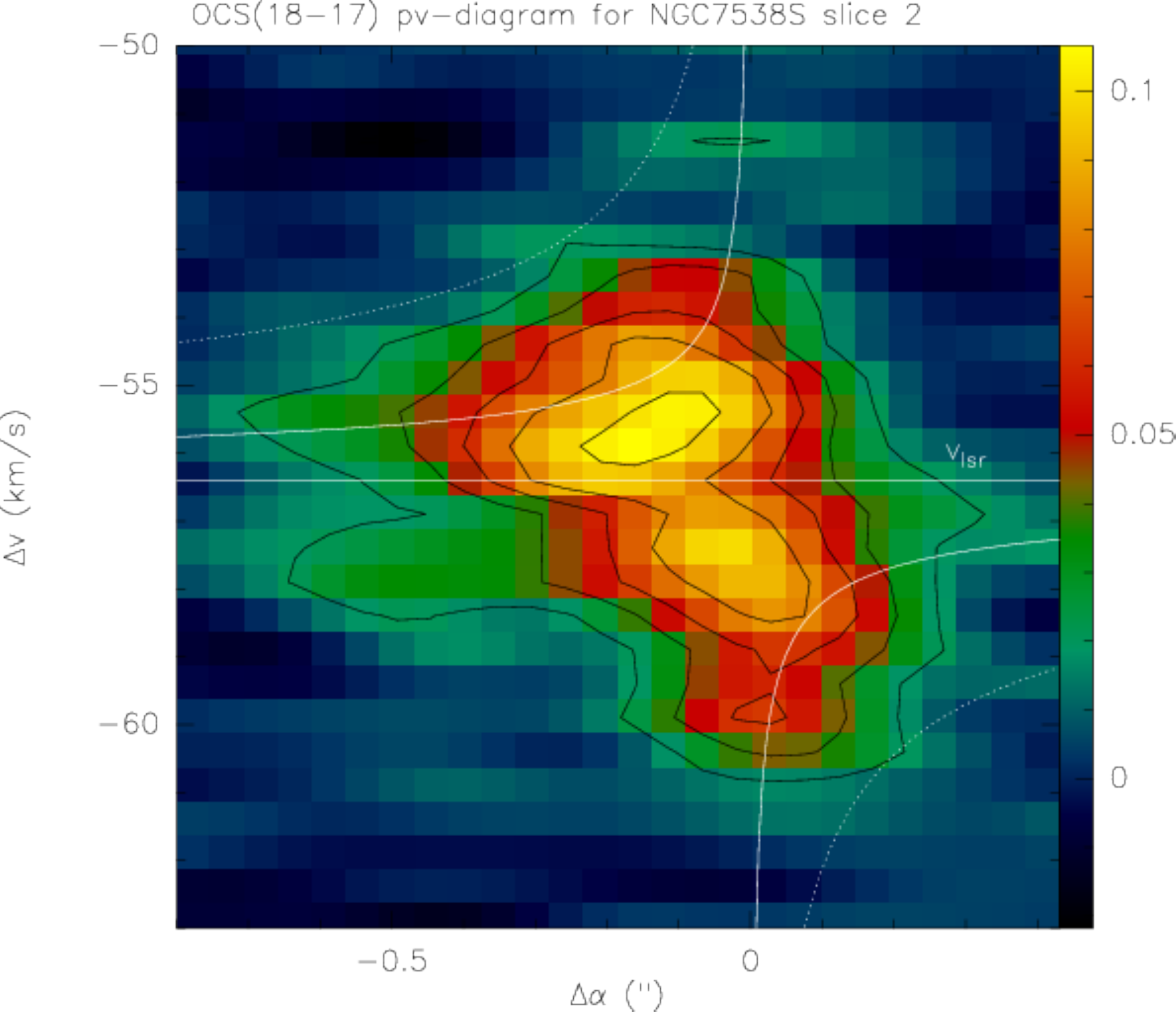}
\includegraphics[width=9cm]{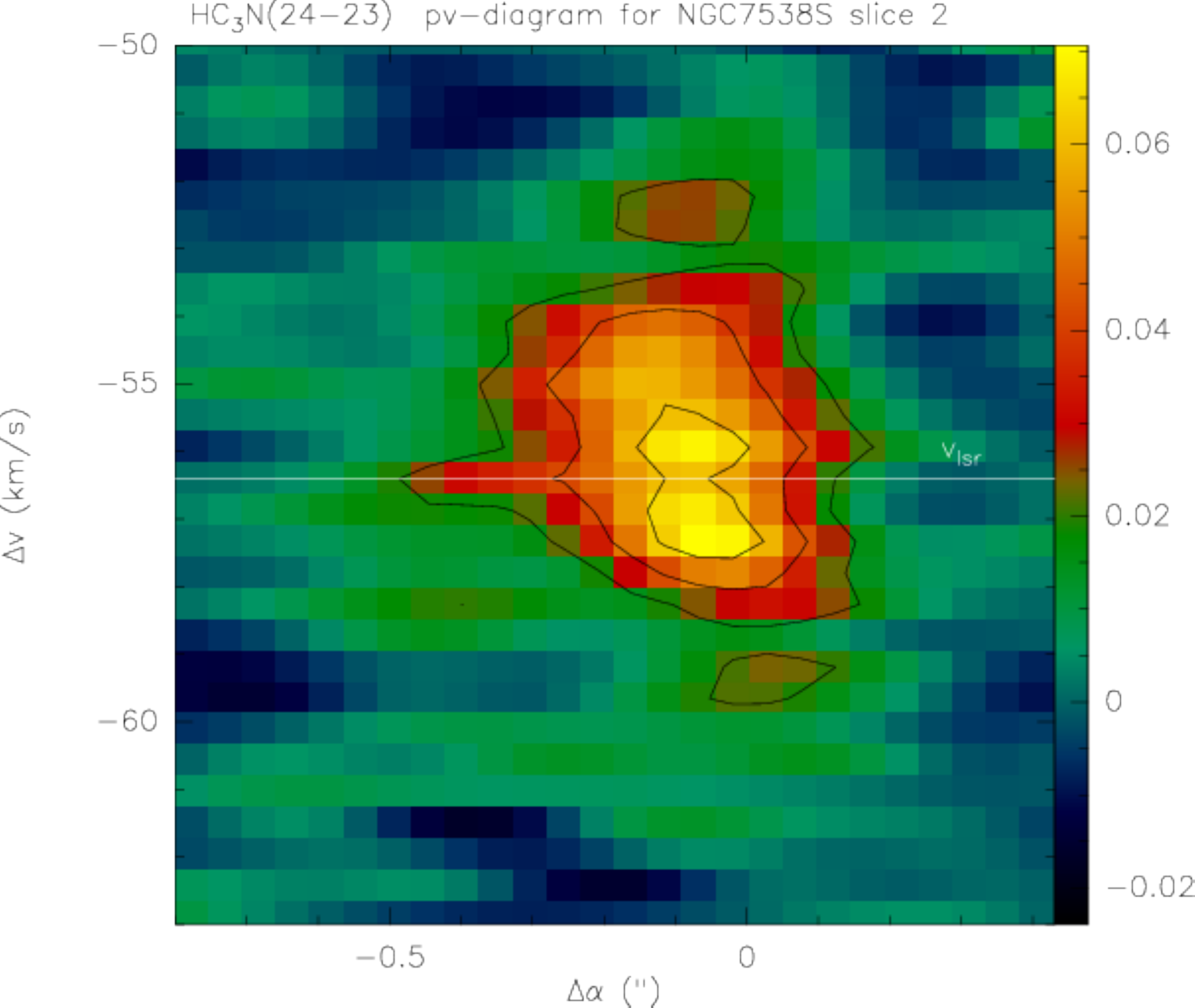}
\includegraphics[width=9cm]{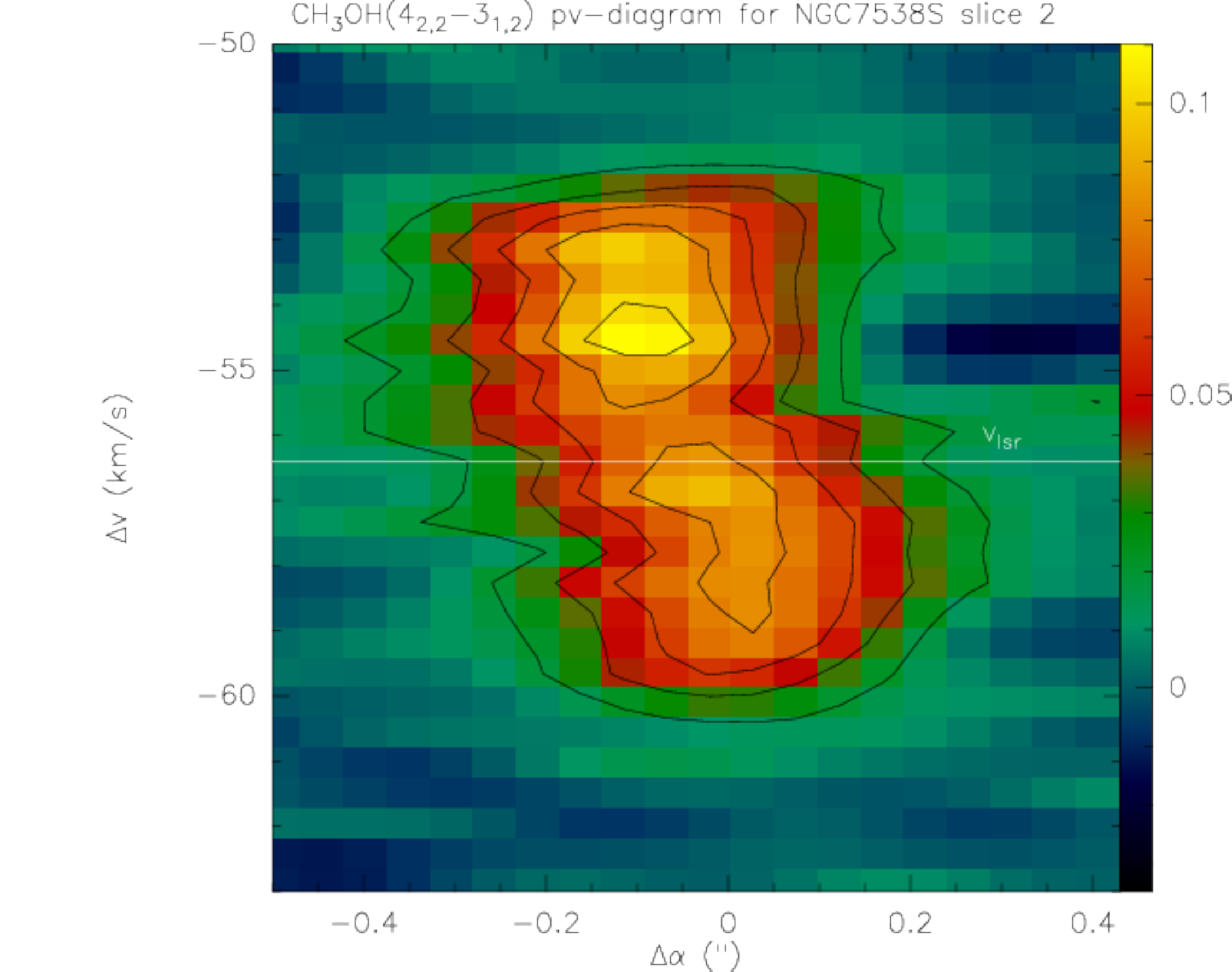}
\caption{Position velocity cuts through NGC7538S mm2 in several lines
  as marked in each panel along the axis shown in
  Fig.~\ref{7538s_overlay}, respectively. The $v_{\rm{lsr}}$ is
  marked. The full and dotted lines in the OCS panel correspond to
  Keplerian rotation curves around a 1 and 10\,M$_{\odot}$ star,
  respectively. The units of the wedges are Jy beam$^{-1}$.}
\label{7538s_mm2_pv}
\end{figure*}

Figure \ref{7538s_overlay} presents the 1st and 2nd moment maps
(intensity weighted peak velocities and line widths) of respective
lines toward that region. While the lower excited lines like
H$_2$CO$(3_{2,2}-2_{2,1})$ or CH$_3$OH$(4_{2,2}-3_{1,2})$ show also a
bit more extended emission, molecular emission from high-density
tracers like CH$_3$CN or HC$_3$N are largely confined to mm1, mm2 and
their close environment. While there is no obvious velocity difference
between mm1 and mm2, toward both sub-peaks we detect in all lines
velocity gradients across the mm continuum peaks, for mm1 almost
north-south and for mm2 in northeast-southwest direction. Toward mm1
it is interesting to note that the molecular emission to the north is
confined almost to the same region as the mm continuum emission
whereas the molecular line data extend significantly outside the
southern $4\sigma$ contours of the mm continuum emission. Since the
main mm1a peak is close to the northern edge of that structure, the
velocity structure of the gas is rather asymmetric with respect to
that peak. The most blue-shifted gas almost peaks toward mm1a and the
further one goes south, the more redshifted the gas gets. For the
higher density lines of CH$_3$CN$(12_k-11_k)$, OCS(18--17) and
HC$_3$N(24--23), the line-width or 2nd moment peak is also not exactly
toward the main mm continuum peak but a little bit to the south,
almost at the tip of the central, north-south elongated mm continuum
contour. As discussed in section \ref{continuum}, we cannot properly
resolve a secondary component there, nevertheless, it appears likely
that this slightly elongated structure will resolve into a binary
system at even higher spatial resolution.

Figure \ref{7538s_mm1_pv1} presents position velocity cuts of selected
lines along the axis marked for mm1 in Figure \ref{7538s_overlay}.
While the emission appears relatively symmetric around the
$v_{\rm{lsr}}$, as already mentioned above, the velocity is not
symmetric around the main mm peak mm1a which is put at offset 0. Even
if one shifts the center by $\sim 0.25''$ south toward the peak of the
2nd moment maps, it still does not appear as a symmetric position
velocity cut. The data clearly confirm that the most blue-shifted gas
is centered on mm1a and the red-shifted emission continuously moves to
the south.  Furthermore, the pv-diagrams do not exhibit any signature
of Keplerian rotation. These signatures indicate that the observed
velocity structure from mm1 unlikely stems from rotation. Since the
jet axis is aligned approximately in northwest-southeast direction
(Figure \ref{cont}), not much offset from the main velocity
gradient observed here, it may well be that the velocity gradient is
strongly influenced by the central jet and outflow.


In comparison to mm1, Figure \ref{7538s_mm2_pv} shows the
position-velocity cuts through mm2 along the axis shown in Figure
\ref{7538s_overlay}. Since mm2 is much smaller in spatial extend and
only barely resolved by our observations, the pv-diagrams also exhibit
less prominent signatures of velocity gradients. Nevertheless, a
velocity gradient is identifiable, most prominently in the OCS(18--17)
line. While the structure is too small for a more detailed analysis,
it is interesting to note that the OCS(18--17) data are at least
consistent with Keplerian rotation around a $\sim$1\,M$_{\odot}$
central object, whereas the expected rotation curve of a more massive
object represents the data significantly worse.  While one should not
take these masses at face-value, they nevertheless indicate that the
central mass in NGC7538S mm2 is significantly lower than that in
NGC7538IRS1.

\begin{figure}[htb] 
\includegraphics[width=9cm]{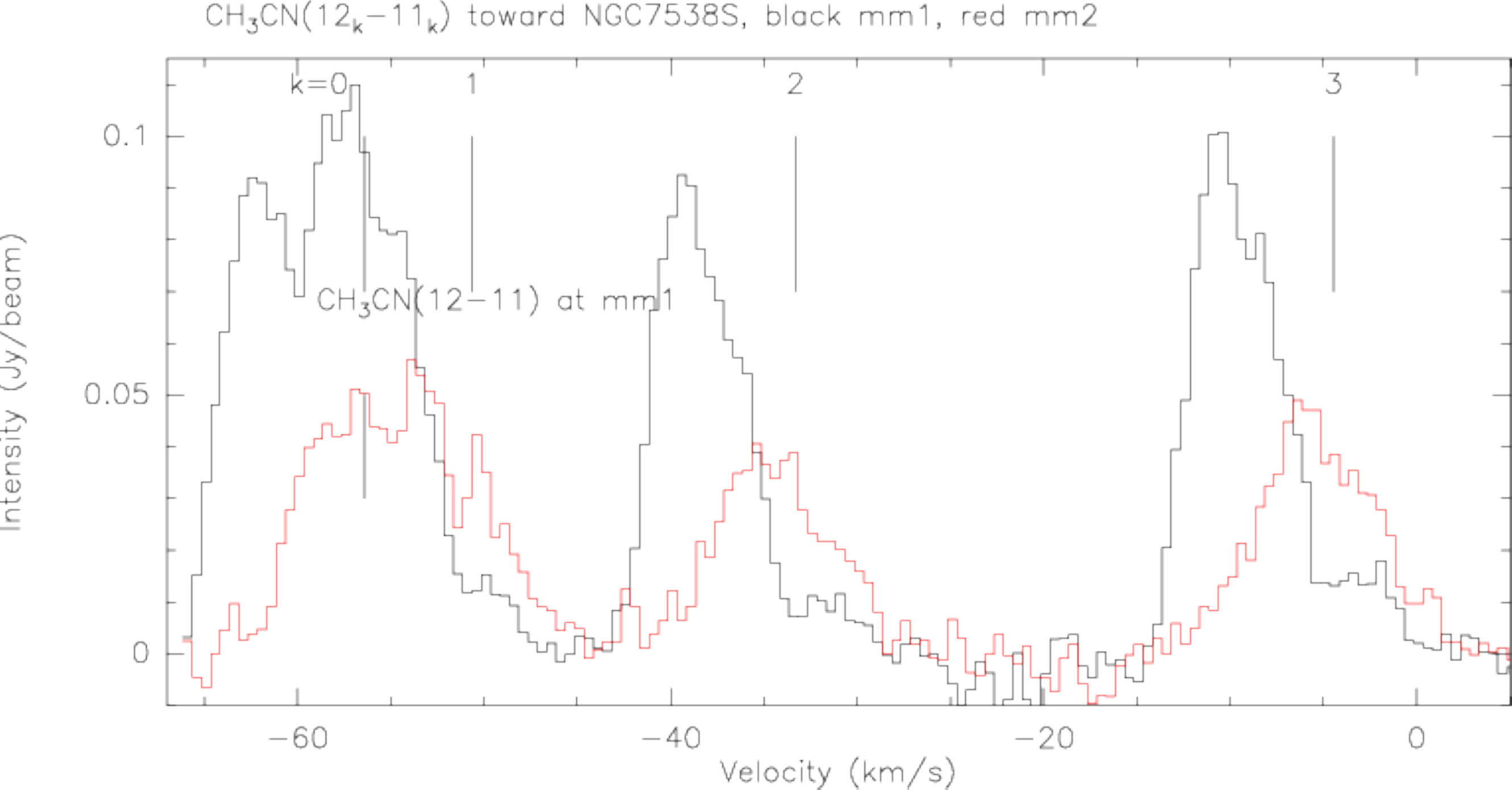}
\caption{CH$_3$CN$(12_k-11_k)$ spectra ($0\leq k\leq 3$) toward
  NGC7538S mm1a (black) and mm2 (red).}
\label{7538s_ch3cn0123}
\end{figure}

\begin{figure}[htb] 
\includegraphics[width=9cm]{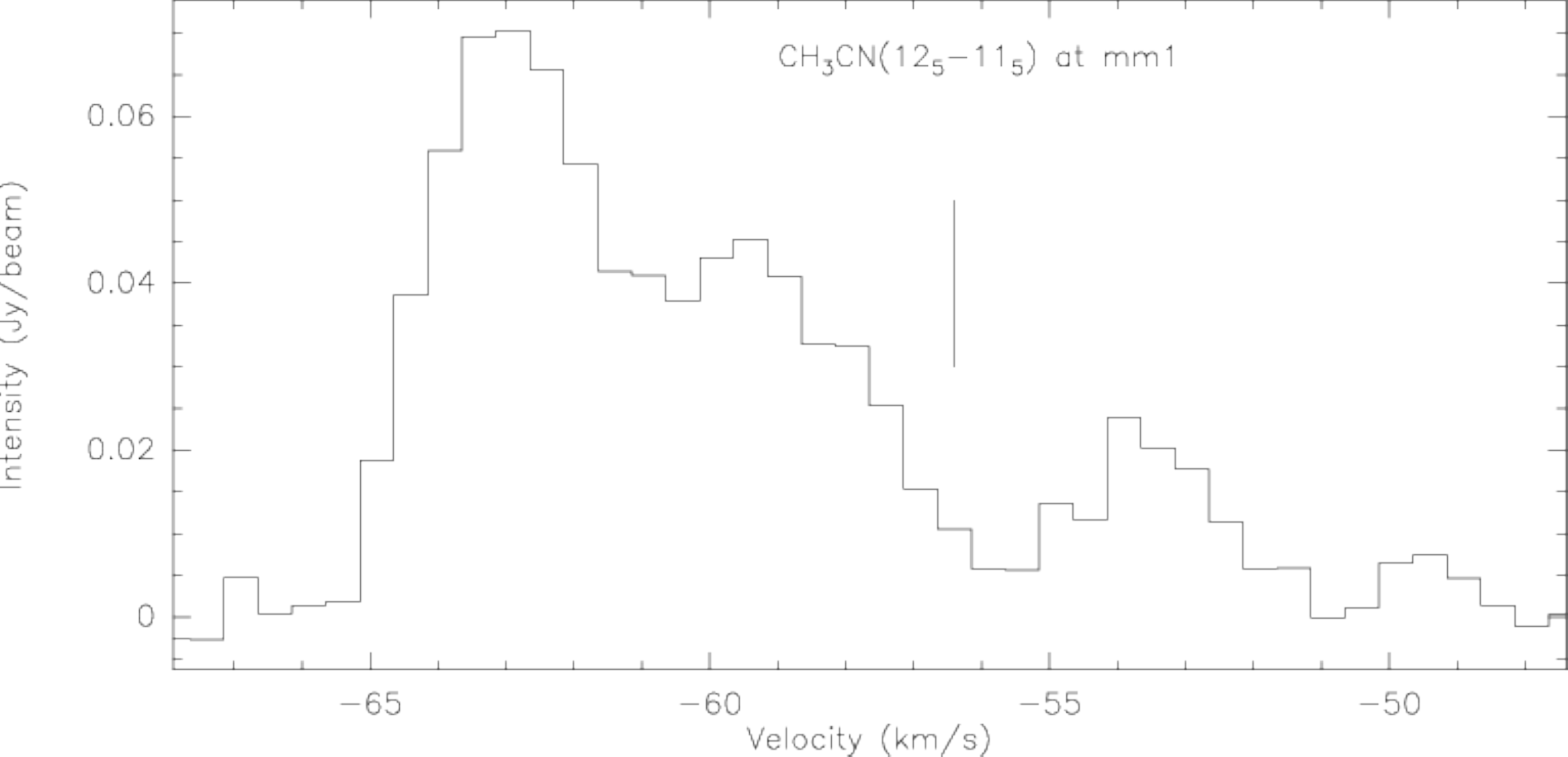}
\includegraphics[width=9cm]{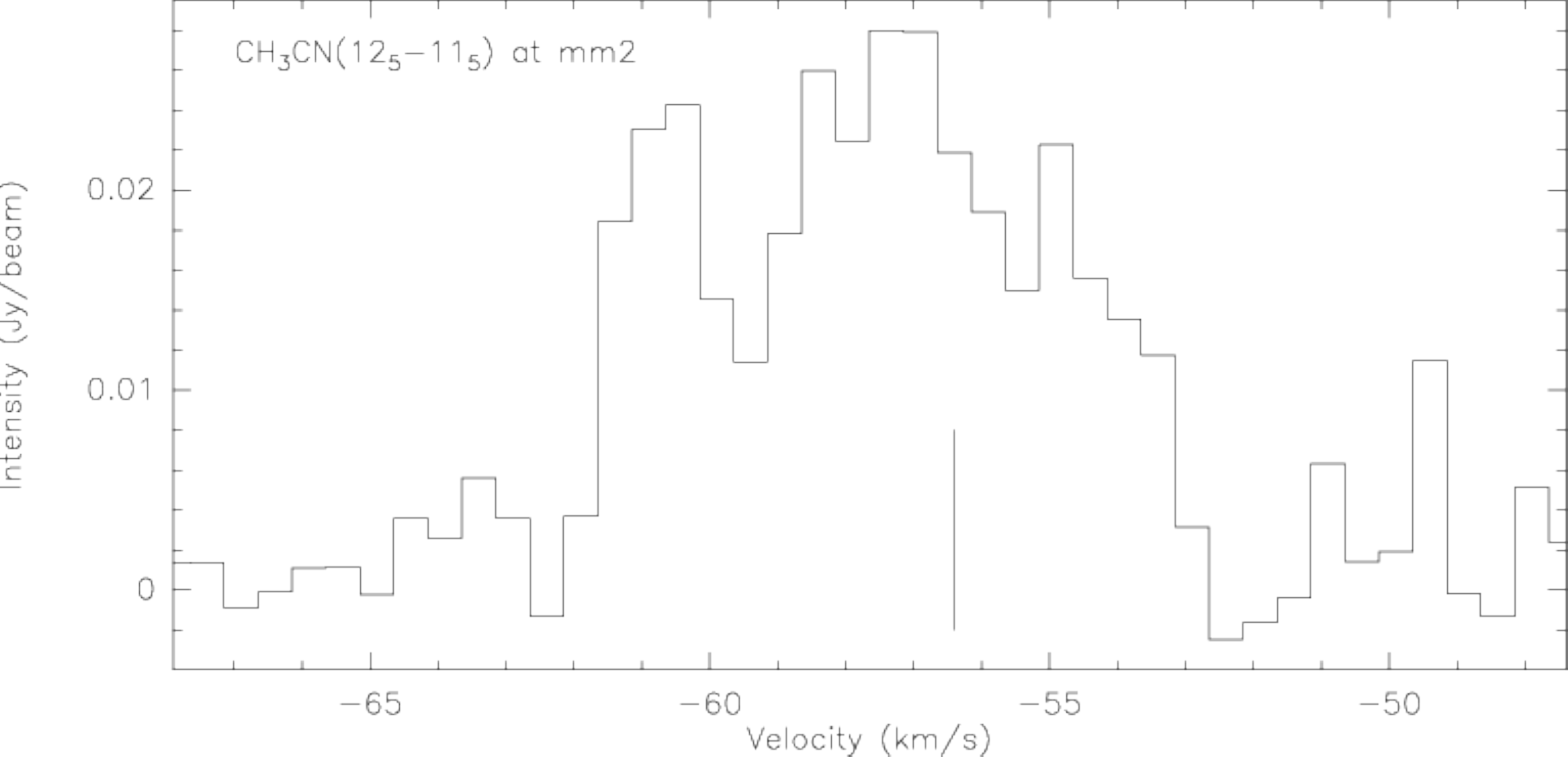}
\includegraphics[width=9cm]{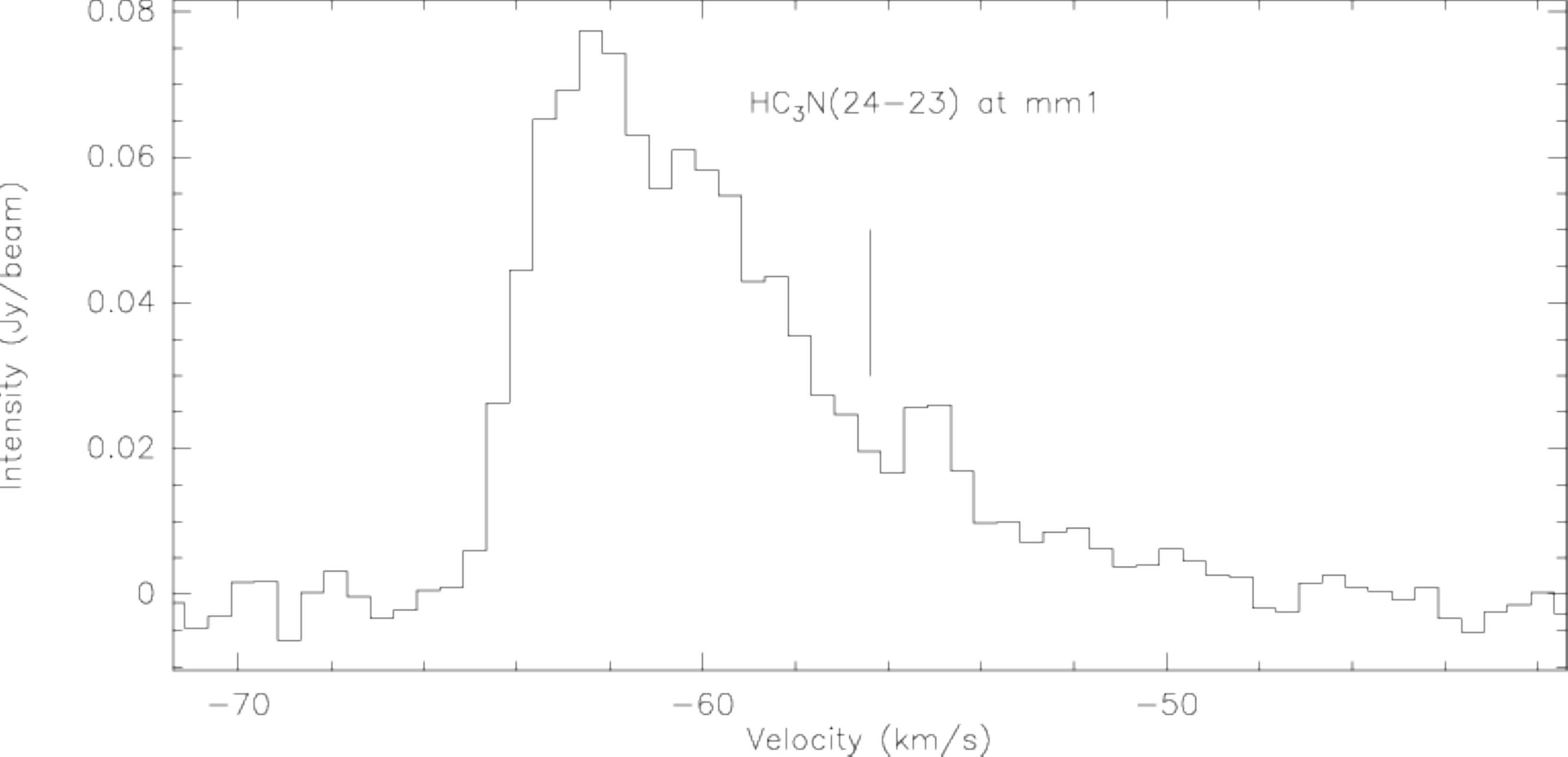}
\includegraphics[width=9cm]{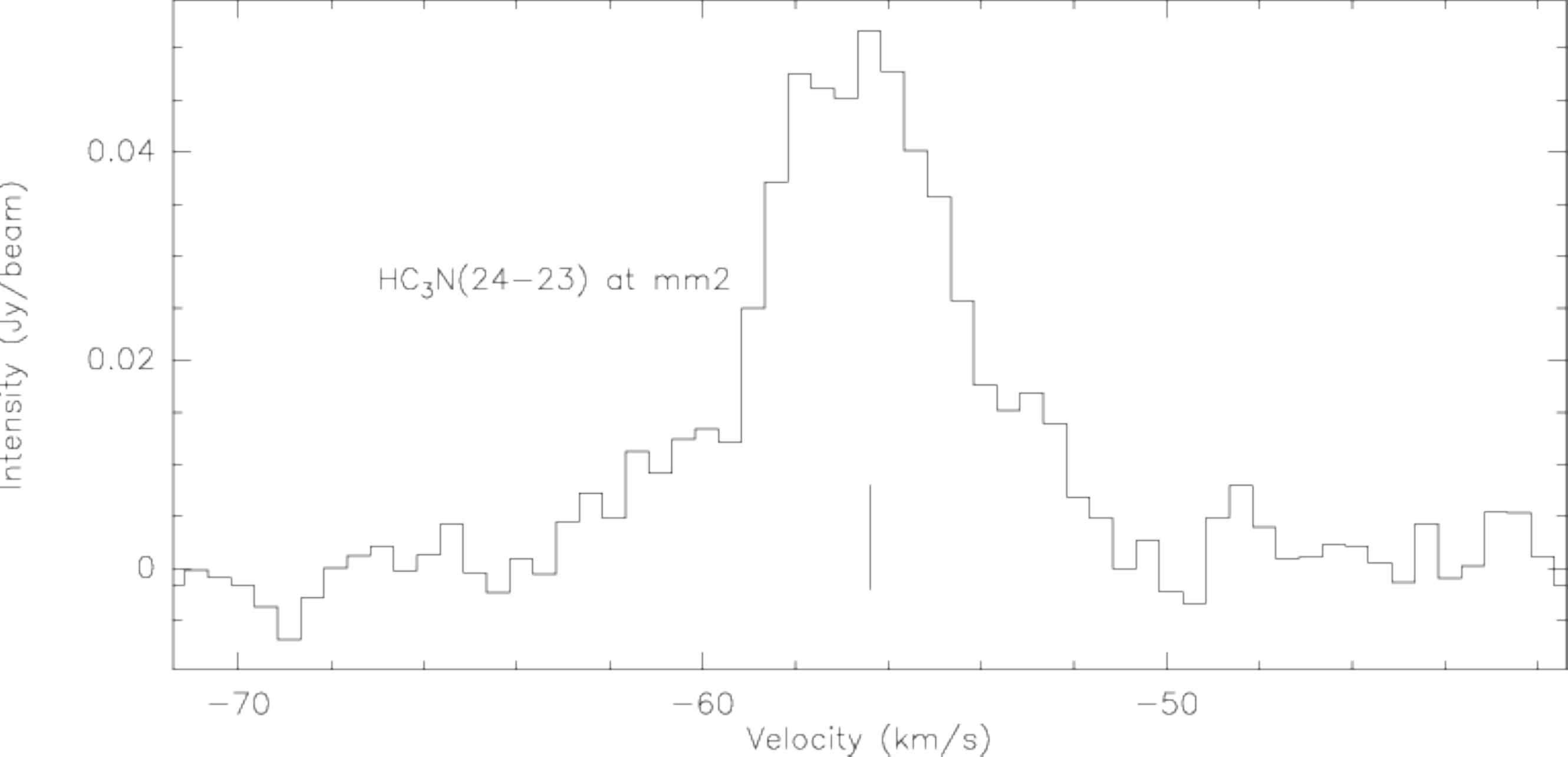}
\caption{CH$_3$CN$(12_5-11_5)$ and HC$_3$N$(24-23)$ spectra toward
  NGC7538S mm1a and mm2.}
\label{7538s_spec}
\end{figure}

A different way to investigate the spectral structure of the two
sources is again via directly looking at the spectra toward the peak
positions (Figures \ref{7538s_ch3cn0123} and \ref{7538s_spec}). While
the spectra toward mm2 exhibit more or less Gaussian shapes around the
$v_{\rm{lsr}}$, this is not the case for the spectra extracted toward
mm1a. The mm1a spectra are strongly dominated by the blue-shifted gas
which was already identified in pv-diagrams, but we see an additional
red-shifted component that is separated by a little flux-depression
(not absorption) around the $v_{\rm{lsr}}$. This will be discussed in
more detail in section \ref{kinematics_ngc7538s}.

\section{Discussion}
\label{discussion}

\subsection{Continuum structure}

\subsubsection{NGC7538IRS1}
\label{cont_discuss_irs1}

While fragmented sub-sources like those in NGC7538S have been observed
regularly with interferometers toward high-mass star-forming regions
(e.g., \citealt{beuther2007c,zhang2009}), the very massive, centrally
condensed and unresolved source NGC7538IRS1 is peculiar and less
typical (e.g., \citealt{bontemps2010}). In particular, the fact that
one observes a near-infrared source toward that position but at the
same time has column densities in excess of $10^{26}$\,cm$^{-2}$
(corresponding to visual extinctions in excess of $10^5$\,mag) is
surprising. This seems to indicate that the outflow/jet from the
source must be very close to the line of sight allowing the infrared
radiation to escape through the outflow cavity. This is reminiscent to
similar sources like W3IRS5 or G9.62+0.19 where also the infrared
sources are detected in spite of high column densities because of the
alignment of the outflow close to the line of sight (e.g.,
\citealt{rodon2008,hofner2001,linz2005}).

Furthermore, it is astonishing how much mass is concentrated in a very
small area in NGC7538IRS1. Table \ref{mm} and Figure \ref{cont} show
that at least more than 40 and potentially even more than
100\,M$_{\odot}$ are concentrated within a source with projected
diameter of approximately 2000\,AU. To the authors knowledge, there is
no other region known with so much mass in such a small area. Assuming
50\,M$_{\odot}$ within a sphere of diameter of $\sim$2000\,AU, this
corresponds to a mean H$_2$ density of $\sim 2.1\times
10^9$\,cm$^{-3}$ which is also high compared to other
star-forming regions. It is also about 2 orders of magnitude larger
than the average densities derived by \citet{qiu2011}. This difference
can largely be attributed to the much smaller size of the core in the
PdBI data (a factor 3 in radius) as well as different assumptions in
the mass calculations (we adopted a slightly different free-free
contribution and gas-to-dust ratio).

\citet{puga2010} report an age spread between 0.5 and 2.2\,Myrs for
the surrounding infrared cluster whereas the various signatures of
ongoing star formation toward the central mm and infrared source
(e.g., maser emission, outflows) indicate that the central and most
massive object is still in an active star formation process. It
appears that in that region the most massive star forms last compared
to the lower-mass population, similar to other studies like, e.g.,
\citet{kumar2006} or \citet{wang2011}. Furthermore, the strong
concentration of mass within a single object without much further
fragmentation (except those on larger scales as reported by
\citealt{qiu2011}, see also \citealt{bontemps2010}) is consistent with
a scaled-up low-mass star formation scenario for the formation of
high-mass stars (e.g.,
\citealt{mckee2002,mckee2003,krumholz2006b,krumholz2009,kuiper2010}).

\subsubsection{NGC7538S}

The substructure one finds in NGC7538S is indicative of hierarchical
fragmentation on different scales. While the single-dish mm continuum
data show only one large-scale gas clump with a projected diameter of
$\sim$0.5\,pc (e.g., Figure \ref{overview},
\citealt{sandell2004,reid2005}), first interferometric observations
already revealed an elongated gas clump with an extend of
$\sim$30000\,AU$\sim$0.15\,pc that showed velocity signatures
indicating rotation. Recently, \citet{wright2012} resolved that
elongated structure into three mm continuum sources which we confirm
here at even higher spatial resolution. Our new data now also indicate
that mm1 splits up likely in $\geq$3\,sources forming a trapezium-like
system (e.g., \citealt{ambartsumian1955,megeath2005,rodon2008}). The
hierarchical fragmentation observed in NGC7538S resembles the
structures recently discussed by \citet{zhang2009} and
\citet{wang2011} for the infrared dark cloud G28.34, although on
projected smaller spatial scales in NGC7538S.

Following the simple toy-model outlined in \citet{beuther2012a}, in a
cluster-forming scenario with a typical \citet{kroupa2001} initial
mass function and a star formation efficiency of $\sim$30\%, one needs
approximately a 1000\,M$_{\odot}$ initial gas clump to form a cluster
with at least one 20\,M$_{\odot}$ high-mass stars. As estimated in
section \ref{continuum}, NGC7538S fulfills that criterium (and
NGC7538IRS1 even more), and while mm2 supposedly does not form a
high-mass star but rather a low- to intermediate-mass object (see
section \ref{lines_ngc7538s}), the higher gas mass (Table \ref{mm})
and the other high-mass star formation indicators discussed in the
introduction and outlined in Figure \ref{cont} indicate that mm1
likely forms a high-mass star within this still very young
cluster-forming region.

\subsection{Kinematics}

\subsubsection{NGC7538IRS1}
\label{kinematics_ngc7538irs1}

The exceptional red- and blue-shifted absorption features presented in
section \ref{lines_ngc7538irs1} give various insights into the
physical processes around that source. As discussed in section
\ref{cont_discuss_irs1}, the jet/outflow from this source has to be
oriented approximately along the line of sight. Therefore,
blue-shifted absorption features against the continuum peak position
should be associated with expanding motions from the jet/outflow. Also
the fact that we observe in the line with the lowest excitation
temperature (H$_2$CO$(3_{0,3}-2_{0,1})$ with $E_u/k \sim 21$\,K, see
also Table \ref{linelist}) an additional absorption feature at even
more blue-shifted velocities is consistent with this picture. The
lower-excited line traces colder gas further away from the source, and
many outflows are known to exhibit Hubble-like velocity structure
where the velocity increases with distance from the source (e.g.,
\citealt{arce2007}). Hence colder gas further outside should show
absorption at velocities further blue-shifted than closer to the
source.

In contrast to that, the red-shifted absorption is indicative of
infalling gas. Following the approach outlined in \citet{qiu2011},
assuming a spherical infall geometry one can estimate mass infall
rates\footnote{Since one does not know whether the gas gets actually
  accreted or not, we prefer the term ``infall rate'' rather than
  ``accretion rate''.} according to $\dot{M}_{\rm{in}} = 4\pi r^2 \rho
v_{\rm{in}}$ where $\dot{M}_{\rm{in}}$ and $v_{\rm{in}}$ are the
infall rate and infall velocity, and $r$ and $\rho$ the core radius
and density. The latter two values are $\sim 1000$\,AU and $\sim
2.1\times 10^9$\,cm$^{-3}$, respectively (see section
\ref{cont_discuss_irs1}). As infall velocity we use 2.3\,km\,s$^{-1}$
which corresponds to the difference between the peak of the
red-shifted absorption at $\sim -55$\,km\,s$^{-1}$ and the
$v_{\rm{lsr}}$ at around $\sim -57.3$\,km\,s$^{-1}$ (Gerner et al.~in
prep, \citealt{vandertak2000}). With these numbers, we derive an infall
rate estimate of $\dot{M}_{\rm{in}} \sim 7\times
10^{-2}$\,M$_{\odot}$\,yr$^{-1}$. This is approximately a factor 20
larger than the rate estimated by \citet{qiu2011}. This difference is
due to the higher spatial resolution of our data where we resolve the
core on smaller scales (radius of 1000\,AU here compared to 3000\,AU
in \citealt{qiu2011}) which additionally results in higher average
densities of the central core (see section \ref{cont_discuss_irs1}).
Considering that the accretion does not occur in a spherical mode over
$4\pi$ but rather along a flattened disk structure with a solid angle
of $\Omega$, the actual disk infall rates $\dot{M}_{\rm{disk,in}}$
should scale like $\dot{M}_{\rm{disk,in}}=\frac{\Omega}{4\pi}\times
\dot{M}_{\rm{in}}$. Based on the simulations by \citet{kuiper2012} and
R.~Kuiper (priv.~comm.), such outflow covers approximately 120\,degree
opening angle and the disk 60\,degree (to be doubled for the
north-south symmetry). Since the opening angle does not scale linearly
with the surface element, full integration results in $\sim$50\% or
$\sim 2\pi$ of the sphere being covered by the disk.  This results in
disk infall rates of $\dot{M}_{\rm{disk,in}}\sim 3.5\times 10^{-2}$
\,M$_{\odot}$\,yr$^{-1}$, still very high and in the regime of
accretion rates required to form high-mass stars (e.g.,
\citealt{wolfire1987,mckee2003}).
 Although we cannot prove that the gas falls in
that far that it can be accreted onto the star (and does not get
reverted by the innermost radiation and outflow pressure), such high
infall rates should be a pre-requisite to allow accretion even when
the central high-mass star has ignited already (e.g.,
\citealt{keto2003,kuiper2010,kuiper2011}).

An additional caveat arises from the potential contribution of the
accretion luminosity to the total luminosity of the region. If one
used the infall rates as actual accretion rates $\dot{M}_{\rm{acc}}$
and estimated the accretion luminosity $L_{\rm{acc}}$ via the
classical $L_{\rm{acc}}= \frac{GM_*\dot{M}_{\rm{acc}}}{R_*}$ (with $G$
the gravitational constant, $M_*$ and $R_*$ the estimated stellar mass
of 30\,M$_{\odot}$ and a stellar radius following
\citealt{hosokawa2009}), one would derive unreasonably high accretion
luminosities in excess of the measured luminosity. Therefore, some
parameters in this equation have to be different. Most likely this is
the accretion rate because not all gas will fall on the star but a
large fraction will likely be expelled again by the energetic outflow.
Estimating that ratio is out of the scope of this paper. Nevertheless,
the data indicate that a significant fraction of the measured
luminosity may still stem from the accretion processes.

As outlined in the introduction, the disk orientation in this region
has been subject to intense discussion. Although the 1st moment map of
NGC7538IRS1 is distorted from the absorption toward the peak, our data
clearly support the orientation of the disk along a
northeast-southwest orientation that was also proposed by
\citet{debuizer2005c}, \citet{klaassen2009} or \citet{surcis2011}. We
do not find signatures of rotational motion along the more east-west
oriented structure that was proposed as a disk from CH$_3$OH maser
observations \citep{pestalozzi2004,pestalozzi2009}. However, we cannot
exclude that the dense gas northeast-southwest oriented emission and
the east-west CH$_3$OH masers belong to the same torus-disk structure
as suggested by \citet {surcis2011}.

\subsubsection{NGC7538S}
\label{kinematics_ngc7538s}

The spectral line signatures in NGC7538S vary considerably among the
three main sub-sources. While mm1 and mm2 are strong line emitters,
mm3 shows no line emission in any of the discussed lines of this
project.  Contrary to that discrepancy, the continuum emission from
mm2 and mm3 is very similar in size, column density and mass.
Therefore, it is most likely that the spectral line differences are
real chemical differences among the two sub-sources mm2 and mm3. It is
tempting to interprete that these differences are due to different
evolutionary stages.  While mm2 shows also rotational signatures and
is already a star-forming core, mm3 may well still be in a younger and
starless phase. In this picture, sources that are separated by less
than 10000\,AU and that are embedded within the same large-scale gas
clump may not evolve coeval at all. What are the physical reasons for
this evolutionary differences? Unfortunately, our data do not allow us
to draw conclusions on that point.

The additional blue-redshifted gas components visible in the spectra
toward mm1a (Figures \ref{7538s_ch3cn0123} and \ref{7538s_spec})
indicate that while the pv-diagrams (Figure \ref{7538s_mm1_pv1}) are
dominated by the larger structure encompassing also the dust
elongation toward the south, there exist additional velocity structure
toward the peak mm1a.  Although we do not spatially resolve that
substructure, it may well stem from a smaller embedded disk centered
on mm1a which is also the likely driving source of the jet in that
region (Fig.~\ref{cont}). Since the continuum source mm1 is already
resolved in at least two sub-sources (mm1a and mm1b), and mm1a is
elongated indicating the existence of an additional source, it is
likely that mm1 hosts a multiple system where large-scale kinematic
structures are present (best visible in the pv-diagram in Figure
\ref{7538s_mm1_pv1}) as well as potential small-scale rotational
structure around individual sub-sources only identified in the spectra
(Figures \ref{7538s_ch3cn0123} and \ref{7538s_spec}). Comparing the
larger almost north-south velocity gradient in mm1 with the
orientation of the jet in approximately northwest-southeast direction,
it appears that the dense gas kinematics in mm1 are strongly
influenced by the jet. This makes the identification of rotational
signatures even harder. It is likely that higher spatial resolution as
well as spectral lines sensitive only to the innermost region around
the central protostar are needed to disentangle the rotational
structure around mm1a from the kinematic signatures caused by the jet.

Going to larger spatial scales, \citet{sandell2003} already identified
a velocity gradient across the whole 30000\,AU structure that
encompass the three regions mm1 to mm3 approximately along the
connecting axis of the sub-sources. Interestingly, \citet{sandell2003}
find that the large-scale rotation is consistent with Keplerian
motion. Since the kinematics around mm1 appear to be dominated by the
jet, Keplerian signatures cannot be expected for that subregion.  The
situation is less obvious for mm2 (Figure \ref{7538s_mm2_pv}) where
the velocity structure at least does not disagree with Keplerian
rotation. Regarding the alignment of axis, the proposed rotational
axis of the large-scale toroid \citep{sandell2003} and the structure
around mm2 are approximately aligned whereas the jet-axis dominating
mm1 is almost perpendicular to that. Unfortunately, such low-number
statistics do not allow us to derive further conclusions from that.

\section{Conclusions}
\label{conclusion}

Very-high-resolution mm continuum and spectral line observations of
the two high-mass disk candidates NGC7538IRS1 and NGC7538S reveal
intriguing information about the small-scale morphology and kinematics of
these two regions. 

NGC7538S appears as a relatively typical source that fragments down to
the smallest resolvable scales. The large-scale single-dish gas clump
forms an elongated torus of $\sim$30000\,AU \citep{sandell2003} that
fragments into three cores with separations on the order of 5000\,AU.
At even higher spatial resolution, these cores show additional
substructure and the most massive one fragments even further. These
data are consistent with hierarchical fragmentation. While the
kinematics of the main mm peak mm1 appears to be strongly influenced
by the jet/outflow emanating from the source, a spectrum extracted
toward the central peak mm1a is indicative of additional unresolved
rotational motions. Higher-resolution data are needed to resolve that.
The spectral lines toward mm2 also exhibit a velocity gradient, and
although barely resolved, the data are consistent with Keplerian
rotation around a low- to intermediate-mass object. Therefore, in
NGC7538S we are witnessing the formation of a very young cluster where
the sources within mm1 have the potential to form a high-mass star at
the end of the evolution. An additional interesting feature is that
mm1 and mm2 are strong spectral line emitters whereas mm3 is not. While
mm2 and mm3 appear very similar in the continuum emission, the strong
diversity in the spectral lines indicate different evolutionary
stages. Hence, even within areas of $\sim$10000\,AU diameter, we find
cores that are likely not evolving coeval. Determining physical
reasons for such discrepancies is beyond the scope of this paper.

NGC7538IRS1 remains a single source even at $\sim$800\,AU spatial
resolution ($\sim 0.3''$). This is even more surprising considering
that the source is embedded in an already existing near-infrared
cluster.  NGC7538IRS1 has extremely large gas and dust column
densities corresponding to visual extinction values on the order of
$\sim 10^5$\,mag. The fact that we still see the central source in the
infrared implies that the jet/outflow from that region should be
aligned closely to the line of sight allowing us to glimpse through
the outflow cavity close onto the central source. The central 2000\,AU
around the source contain a large gas mass on the order of
50\,M$_{\odot}$, implying central average densities in the regime of
$10^9$\,cm$^{-3}$.

Since the position-velocity diagrams of NGC7538IRS1 are distorted by
the absorption, interpretation of kinematic signatures are more
difficult. Nevertheless, we clearly identify a velocity gradient in
northeast-southwestern direction, consistent with the proposed
mid-infrared disk emission orientation \citep{debuizer2005c} and
perpendicular to the outflow axis. Our data do not support the
proposed rotational axis based on CH$_3$OH maser emission
\citep{pestalozzi2004,pestalozzi2009} that is inclined to our observed
axis by approximately 60\,deg.

At $\sim$0.3$''$ spatial resolution, almost all observed spectral lines
reveal strong absorption signatures toward the peak of the mm
continuum emission (that coincides within the errors with IRS1) in
NGC7538IRS1. While some lines, in particular the lower excitation
temperature lines like those of H$_2$CO appear to be dominated by
blue-shifted absorption indicative of outflowing gas, the
higher-excitation and higher-density lines exhibit clear red-shifted
absorption that has to be due to infalling gas. Since the jet/outflow
is supposed to be aligned along the line-of-sight, it is no surprise
that infalling and outflowing gas are observed at the same spatial
position. Estimated mass infall rates are very high, on the order of
$10^{-2}$\,M$_{\odot}$\,yr$^{-1}$. Although we cannot proof that the
gas will continue to be accreted by the central star, nevertheless,
the conditions are sufficient to allow accretion still during that
already luminous and active phase of the protostellar evolution.
Combining the large infall rates with the fact of barely any
fragmentation of the gas and dust core, these data are consistent with
high-mass star formation proceeding in a scaled-up version of low-mass
star formation.

While the presented data already reveal many new insights for both
regions, significant information is still missing. In particular,
the proposed accretion disk signatures for both sources --
NGC7538IRS1 and NGC7538S mm1 -- are ``contaminated'' by absorption and
jet signatures, respectively. To overcome these issues, one likely
needs to resort to even higher excited lines that are neither absorbed
by the envelope nor emitted by the outflowing gas.

\begin{acknowledgements} 
  We like to thank the IRAM staff, in particular Jan Martin Winters,
  for all support during the observation and data reduction process.
  Thanks a lot also to Goeran Sandell for providing the large-scale
  1.2\,mm continuum map presented in Figure 1.  Furthermore, we
  thank the referee Eric Keto for his positive and constructive
  report on tha paper.
\end{acknowledgements}


\end{document}